\documentclass{llncs}

\usepackage{hyperref}
\usepackage{proof}
\usepackage{amssymb}
\usepackage{amsmath}
\usepackage{tikz}
\usetikzlibrary{shapes}
\usepackage{xspace}
\usepackage{listings}
\usepackage{graphicx}

\definecolor{codegreen}{rgb}{0,0.6,0}
\definecolor{codegray}{rgb}{0.5,0.5,0.5}
\definecolor{codepurple}{rgb}{0.58,0,0.82}
\definecolor{backcolour}{rgb}{0.95,0.95,0.92}

\lstdefinelanguage{AnB}
{
    morekeywords={Protocol, Types, Actions, Knowledge, Goals},
    sensitive=true,
    morecomment=[1]{\#}, 
}

\lstdefinestyle{mystyle}{
    backgroundcolor=\color{backcolour},   
    commentstyle=\color{codegreen},
    keywordstyle=\color{magenta},
    keywordstyle=[2]\color{blue},
    numberstyle=\tiny\color{codegray},
    stringstyle=\color{codepurple},
    basicstyle=\ttfamily\footnotesize,
    breakatwhitespace=false,         
    breaklines=true,                 
    captionpos=b,                    
    keepspaces=true,                 
    numbers=left,                    
    numbersep=5pt,                  
    showspaces=false,                
    showstringspaces=false,
    showtabs=false,                  
    tabsize=2,
}

\lstnewenvironment{code}
    {\lstset{}%
      \csname lst@SetFirstLabel\endcsname}
    {\csname lst@SaveFirstLabel\endcsname}
\title{PFMC: A Parallel Symbolic Model Checker for Security Protocol Verification}
\author{Alex James \and Alwen Tiu \and Nisansala Yatapanage }
\institute{School of Computing, The Australian National University
}

\begin{document}

\maketitle

\begin{abstract}
We present an investigation into the design and implementation of a parallel 
model checker for security protocol verification that is based on a symbolic model of the adversary,
 where instantiations of concrete terms and messages are avoided until needed to resolve a particular
  assertion. 
We propose to build on this naturally lazy approach to parallelise this symbolic state exploration and
 evaluation.  We utilise the concept of \textit{strategies} in Haskell, which abstracts away
  from the low-level details of thread management
and modularly adds parallel evaluation
strategies (encapsulated as a monad in Haskell). We build on an existing symbolic model checker,
OFMC, which is already implemented in Haskell. We show that there is a very significant speed up of around
 3-5 times improvement when moving from the original single-threaded implementation of OFMC to our multi-threaded 
 version, for both the Dolev-Yao attacker model and more general algebraic attacker models. We identify
  several issues in parallelising the model checker: among others, controlling growth of memory consumption,
   balancing lazy vs strict evaluation, and achieving an optimal granularity of parallelism. 
\end{abstract}

\section{Introduction}

A security protocol describes a sequence of actions and message exchanges between communicating partners in a networked system, in order to achieve certain security goals, such as authentication and secrecy. 
The analysis of security protocols against a stated security property is a challenging problem, not just from a computational complexity perspective (e.g., the problem of establishing the secrecy property of a protocol is undecidable in general~\cite{Durgin99}), 
but also from a protocol designer perspective, since proofs of security properties are dependent on the adversary model, which can be challenging to precisely formalise.  

A commonly used adversary model is the Dolev-Yao model~\cite{DolevY83}. The Dolev-Yao model
assumes that the attacker controls the network, and hence will be able to intercept, modify and 
remove messages. However, the attacker is also assumed to be unable to break the basic cryptographic
primitives used in the protocol. For example, an encrypted message can only be decrypted
by the attacker if and only if the attacker possesses the decryption key. The Dolev-Yao model may be further extended by adding various abstract algebraic properties (e.g., theories for modelling exclusive-or, or Abelian groups in general). In the context of protocol analysis, the Dolev-Yao model and/or its algebraic extensions are sometimes referred to as the symbolic model. 

In this paper, we are concerned with verifying protocols with a bounded number of sessions
in the symbolic model, and we restrict to only proving reachability properties, i.e., those properties that can be expressed as a predicate on a single trace of a protocol run, such as secrecy and authentication. 
Bounded verification aims primarily at finding attacks, but even for a small number of sessions, the complexity of finding attacks is still very high, e.g., NP-complete for the standard Dolev-Yao model~\cite{RusinowitchT01}. 
Another interesting use of bounded verification is to lift the result of such verification to the unbounded case, for a certain class of protocols~\cite{ArapinisDK08}. 
A related work by Kanovich et al. \cite{Kanovich14} on a bounded-memory adversary (which implies bounded number of sessions) also points to  the fact that many attacks on 
existing protocols can be explained under a bounded-memory adversary. These suggest that improving the performance of bounded protocol verifiers may be a worthwhile research direction despite the prevalence of protocol verifiers for unbounded sessions such as Proverif~\cite{Blanchet16} and Tamarin~\cite{MeierSCB13}.

A bounded-session protocol verifier works by state exploration (search), so naturally we would seek to improve its performance by parallelising the search. Given the ubiquity of multicore architecture in modern computing, it is quite surprising that very few protocol verifiers have attempted to benefit from the increase in computing cores to speed up their verification efforts. An important consideration in our attempt is avoiding excessive modification of the (implementation of)  decision procedures underlying these verifiers, as they are typically complex and have been carefully designed and optimised. This motivates us to consider a language where parallelisation (of a deterministic algorithm) can be added as an effect -- Haskell parallelisation monads (e.g., \cite{MarlowNJ11}) fit this requirement very well. In this work, we use OFMC, which is implemented in Haskell, as the basis of our study, looking at ways to introduce parallelisation without changing the underlying implementation of its decision procedures. This will hopefully provide us with a recipe for applying similar parallelisation techniques to other protocol verifiers written in Haskell, notably Tamarin. 

We have implemented a parallel version of OFMC~\cite{ModersheimV09}, which we call PFMC, in Haskell. We show that PFMC significantly improves the performance of OFMC, with a speed-up of around 3-5 times faster than OFMC, when run on 4-16 cores. This promising result allows us to push the boundary of bounded protocol verification. As part of the implementation of PFMC, we have designed what we believe are several novel parallel evaluation strategies for buffered (search) trees in the context of symbolic model checking for protocol verification, which allows us to achieve parallelism with a generally constant memory residency.

{\em Related work.}
Currently, there are not many major protocol verifiers that explicitly support parallelisation. The only ones that we are aware of are the DEEPSEC prover~\cite{Cheval18SP} and the Tamarin prover~\cite{MeierSCB13}. DEEPSEC uses a process-level parallelisation, i.e., each subtask in the search procedure is delegated to a child process and there is no support for a more fine-grained thread-level parallelisation. % as we attempt here. 
Tamarin is implemented in Haskell, which has a high-level modular way of turning a deterministic program into one which can be run in parallel, sometimes called semi-explicit parallelisation. This is the method that we will adopt, as it seems like the most straightforward way to gain peformance with minimal effort. As far as we know, there has been no published work on systematically evaluating the performance of Tamarin parallelisation; some preliminary investigations into its parallelisation performance is given in Appendix~\ref{sec:tamarin}. 
Some of the established protocol verifiers such as Proverif~\cite{Blanchet16}, DEEPSEC~\cite{Cheval18SP}, SATEQUIV~\cite{Cortier18ESORICS}, SPEC~\cite{TiuNH16} or APTE~\cite{Cheval14} are implemented in OCaml, and the support for multicore is not as mature as other languages. %OFMC and 
For an overview of protocol verification tools, see a recent survey article by Barbosa et. al.~\cite{BarbosaBBBCLP19}.

{\em Outline.} The rest of the paper is organised as follows. 
Section~\ref{sec:lts} provides an overview of the state transition system arising from symbolic protocol analysis. Section~\ref{sec:parallel} gives a brief overview of Haskell parallelisation features. Section~\ref{sec:strategies} presents the parallelisation strategies implemented in PFMC and the evaluation of their performance on selected protocol verification problems.   
Section~\ref{sec:conc} concludes and discusses future directions. 
The full source code of PFMC is online.\footnote{\url{https://gitlab.anu.edu.au/U4301469/pfmc-prover}}
\section{Protocol specifications and transition systems}
\label{sec:lts}
There are a variety of approaches for modelling protocols and their transition systems, such as using multiset rewriting~\cite{CervesatoDLMS99}, process calculus~\cite{AbadiG99}, strand spaces~\cite{ThayerHG99}
or first-order Horn clauses~\cite{Blanchet11}. OFMC supports the modelling language IF (Intermediate Format), but it also supports a more user-friendly format known as AnB (for \textit{Alice and Bob}). 
The AnB syntax can be translated to the IF format, for which a formal semantics is given in \cite{AlmousaMV15}; we refer the reader to that paper. 

The operational semantics of OFMC is defined in terms of strand spaces. One can think of a strand as a sequence of steps that an agent takes. The steps could be a sending action, a receiving action, checking for equality between messages and special events (that may be used to prove certain attack predicates).  
We use the notation $A_1 \| \ldots \| A_n$ to denote $n$ parallel strands, $A_1,\ldots,A_n$, that may be associated with some agents. A {\em state} is a triple consisting of a set of strands of honest agents, a strand for the attacker, and a set of events which have occurred. 
The attacker strand consists of the messages the attacker receives by intercepting communication between honest agents, and the messages the attacker synthesises and sends to honest agents. OFMC represents these strands symbolically. For example, the messages the attacker synthesises are initially represented as variables and \textit{concretised} as needed when agents check for equality between messages. 
The attacker strand is represented as a set of deducibility contraints, and the transition relation must preserve satisfiability of these contraints. 

In OFMC, the state transition relation is defined from an adversary-centric view. This means in particular that what matters in the transition is the update to the attacker's knowledge, and the only way to update the attacker's knowledge is through the output actions of the honest agents. Therefore, it makes sense to define a state transition as a combination of input action(s) that trigger an output action from an honest agent, rather than using each individual action to drive the state transition. Due to the presence of parallel composition protocol specifications, the (symbolic) state transitions can generate a large number of possible traces, due to the interleaving of parallel strands. The search procedure for OFMC is essentially an exploration of the search tree induced by this symbolic transition system.

\section{Haskell parallelisation strategies}
\label{sec:parallel}

This section provides a very brief overview of Haskell parallelisation features and discusses some initial unsuccessful attempts (in the sense that we did not achieve meaningful improvements) 
to parallelise OFMC, to motivate our designs in Section~\ref{sec:strategies}. We use the semi-explicit parallelisation supported by Haskell. We assume the reader is familiar with the basics of Haskell programming, and will only briefly explain the parallelisation approach we use here. 

In the semi-explicit parallelism approach in Haskell, the programmer specifies which parts of the code should be paralellised, using primitives such as \lstinline[basicstyle=\small\ttfamily]{par}. The statement \lstinline[basicstyle=\small\ttfamily]{x par y} is semantically identically to \lstinline[basicstyle=\small\ttfamily]{y}. However, the former tells the Haskell runtime system (RTS) to create a worker to evaluate \lstinline[basicstyle=\small\ttfamily]{x} and assigns it to an available thread to execute. In Haskell terminology, such a unit of work is called a {\em spark}. The programmer does not need to explicitly create and destroy threads, or schedule the sparks for execution. The RTS uses a {\em work-stealing} scheduling to execute sparks opportunistically. That is, each available core keeps a queue of sparks to execute and if its queue is empty, it looks for sparks in other cores' queues and `steals' their sparks to execute. This should ensure all available cores are used during runtime. 
An appealing feature of this approach is that one does not need to be concerned with 
low-level details of thread management and scheduling, as they are all handled 
automatically by the Haskell runtime system. 

{\em Determining the granularity of parallelisation.}
There are three main subproblems in the search for attacks that we examined for potential parallelisation: 
\begin{itemize}
\item checking the solvability of a constraint system,
\item enumerating all possible next states from the current state and
\item the overall construction of the search tree itself. 
\end{itemize}
The search for the solutions for a constraint problem can itself be a complex problem, depending on the assumed attacker model. 
Thus it may seem like a good candidate to evaluate in parallel.
However, it turns out that when verifying real-world protocols, most of the constraints 
generated are simple, and easily solvable sequentially. Attempting to parallelise this 
leads to worse performance than executing them sequentially, as it ends up creating too many lightweight sparks. It may be the case that as the number of sessions grows, the constraints generated become larger the deeper down the search tree, so at a deeper node in the search tree, such a parallelisation might be helpful. However, to reach that stage, there would likely be a lot of simple constraints that need to be solved for which the overhead of paralellisation outweighs its benefit. 

Our next attempt was to parallelise the process for enumerating successor states, which involves solving the intruder constraint problem. This led to some improved performance, but was harder to scale up, as it still created too many sparks, many of which ended up being garbage collected, a sign that there were too many useless sparks. 
The final conclusion seems to be that focussing on parallelising the construction of the search tree, executing the constraint solving and the successor state enumeration sequentially, produces the most speed up.

{\em Lazy evaluation and parallelism.}
Haskell by default uses a lazy evaluation strategy in evaluating program expressions. 
By default, functions in Haskell are evaluated to {\em weak head normal form} (WHNF). 
Haskell provides libraries to force a complete evaluation of a program expression, e.g. via the \lstinline[basicstyle=\small\ttfamily]{force} function. This, however, needs to be used with extreme care as it can create unnecessary computation and a potential termination issue. 

One advantage of lazy evaluation, when parallelising a search algorithm, is that it allows one to decouple the search algorithm and the parallelisation strategy. In our case, we can implement the search algorithm as if it is proceeding sequentially, constructing a potentially infinite tree of states, without considering termination, as each node will be evaluated lazily only when needed, i.e. when the attack predicate is evaluated against the states in the search tree. Thus, at a very high level, given a function \lstinline[basicstyle=\small\ttfamily]{f} that constructs a search tree sequentially, and a strategy \lstinline[basicstyle=\small\ttfamily]{s} for parallelisation, the sequential execution of \lstinline[basicstyle=\small\ttfamily]{f} can be turned into a parallel execution using the composition:
\begin{code}
    (withStrategy s) . f 
\end{code}

Applying this strategy to OFMC, 
we applied a custom parallelisation strategy (see Section~\ref{sec:strategy}) to the construction of the
search tree. 
The search algorithm itself does not make any assumption about termination of the search. It does, however, allow the user to specify the depth of search, so any nodes beyond a given depth will not be explored further. 

Profiling the runtime behaviour of the parallelisation strategy revealed that a significant amount of time is spent on garbage collection. When searching at a depth $d$, OFMC keeps the subtrees at depth $> d$, which end up being garbage collected, as they are not needed in the final evaluation of the attack predicate. A solution is to prune the search tree prior to evaluating the predicate on the nodes.
This seems to significantly reduce the memory footprint of the program.

\section{Parallel Strategies for Search Trees}
\label{sec:strategies}

We now present a number of parallelisation strategies implemented in PFMC. The actual implementation contains more experimental parallelisation strategies not covered here, as they did not seem to offer much improvements over the main strategies presented here. 
We note that it is possible to directly benefit from multicore support for the Haskell runtime by compiling the program with the \lstinline[basicstyle=\small\ttfamily]{-threaded} option. However, doing so results in worse performance compared to the single-threaded version, at least in the case for OFMC.

A main difficulty in designing an efficient parallelisation strategy in the case of OFMC is that the branching factor of a given node in the search tree is generally unbounded. Based on our profiling of the search trees of some sample protocols,
the branching factor is highly dependent on the number of sessions of the protocol, the assumed intruder model (e.g., constraint solving for some algebraic theories can yield a variable number of solutions), and the depth of the search tree. This makes it difficult to adapt general parallel strategies for bounded trees such as the ones discussed in \cite{Totoo16}. 

The source code for each strategy discussed here can be found in the file \lstinline[basicstyle=\small\ttfamily]{src/TreeStrategies.hs} in the PFMC distribution.

\vspace{-11pt}

\subsection{parTreeBuffer: a buffered parallel strategy for search trees}
\label{sec:strategy}

A naive parallelisation strategy for OFMC is to simply spark the entire search tree, i.e., for each node in the search tree, we create a spark and evalute it eagerly in parallel. In our tests, it quickly exhausted the available memory in our test server for large benchmark problems, so it does not scale well.
Haskell (through the \lstinline{Control.Parallel.Strategies} library) provides two ways to avoid creating a large number of sparks simultaneously, via a {\em chunking} strategy (\lstinline[basicstyle=\small\ttfamily]{parChunk}), and a {\em buffered} strategy (\lstinline[basicstyle=\small\ttfamily]{parBuffer}). They are however restricted to lists. The chunking strategy, as the name suggests, attempts to partition the input list into chunks of a fixed size, and executes all chunks in parallel, but keeping the sequential execution within each chunk. The buffered strategy attempts to stall the creation of sparks, by first creating $n$ sparks (for a given value $n$), and then \textit{streaming} the sparks one at a time as needed. 

\begin{figure}
    \begin{center}
    \includegraphics[width=0.5\linewidth]{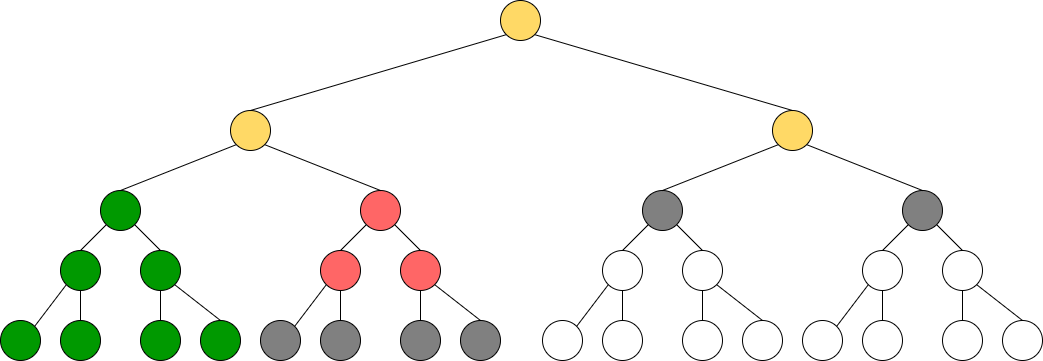}
    \end{center}
    \vspace{-5mm}
    \caption{A buffered search tree.}
    \label{fig:spark3}
\end{figure}

It was not clear what would be an equivalent of \lstinline[basicstyle=\small\ttfamily]{parChunk} applied to potentially unbounded search trees. 
In an initial attempt, we tried to flatten the search tree to a list, use the chunking strategy for lists, 
and `parse' the list back to a tree. This did not produce the desired effect. 
We instead designed an approximation of a buffered strategy, but applied to search trees, by controlling the depth of sparking, which we call {\em par-depth}. 
For example, for a \textit{par-depth} of 2, all nodes at depth less than or equal to 2 will be sparked, and anything beyond depth 2 will be suspended -- by returning their WHNF immediately.  
The sparks created for nodes that have only been evaluated to WHNF will not attempt to create more sparks for the
subtrees. 
When a suspended node is required by the main thread (e.g., when it needs to be evaluated against a security property), it will trigger another round of sparking (up to {\em par-depth})). 
Figure~\ref{fig:spark3} shows a situation where some of the left-most nodes (marked with the green color) have completed their tasks, and the main thread is starting to query the next suspended node (grey node). This triggers sparking of a subtree of depth 2. As can be seen, the strategy essentially sparks a chunk of seven nodes at a time, in a depth-first manner.  
The grey nodes represent those that have been only evaluated to WHNF. The yellow nodes represent nodes that may be either finished performing their task, or waiting for results from their child nodes.

This example shows a balanced binary tree. In reality, the search tree may not be balanced, and the branching factor can vary from one node to another. It may be theoretically possible to create a search tree that is infinitely branching, so our strategy cannot completely guarantee that the number of sparks at any given time would be bounded. 
Indeed, based on our profiling of search trees for a number of benchmarks in OFMC, the search trees can be quite unbalanced, and the branching factors vary greatly between different depths of the search tree. For three sessions of the Kerberos protocol, for example, the branching factors seem to congregate at two extremes: between 1-3 branches at the lower-end versus 40-50 branches at the higher end. In our experiments, we witnessed a relatively constant memory consumption throughout the execution of the benchmark problems up to three sessions. Going beyond four sessions, for large protocols such as Kerberos, the longer the benchmark is executed, the memory consumption may grow substantially, to the point that the entire system memory can be exhausted. Nevertheless, this buffered strategy is simple enough and serves as a good starting point in investigating various trade-offs between the number of cores, the degree of parallelisation and the memory consumption. 

Our buffered strategy is implemented as shown in Figure~\ref{fig:code}. 
The difference between sparking new nodes and stalling the sparking lies in the use of parBuffer vs. parList: the former evaluates a node into WHNF and returns without creating a new spark, whereas the latter would spark the entire list.

\begin{figure}[t]
\begin{code}
    data Tree a = Node a [Tree a]
    parTreeBuffer :: Int -> Int -> Strategy a -> Strategy (Tree a)
    parTreeBuffer _ _ strat (Node x []) = do 
      y <- strat x 
      return (Node y [])
    parTreeBuffer 0 n strat (Node x l) = do 
      y <- strat x
      l' <- parBuffer 50 (parTreeBuffer n n strat) l  
      return (Node y l')
    parTreeBuffer !m n strat (Node x l) = do 
      y <- strat x 
      l' <- parList (parTreeBuffer (m-1) n strat) l   
      return (Node y l')        
\end{code}
\vspace{-2mm}
\caption{A buffered parallel strategy for search trees}
\label{fig:code}
\end{figure}

To evaluate the performance of the \lstinline{parBufferTree} strategy, we selected three benchmark problems: a flawed version of Google Single Sign-On (SSO) protocol~\cite{ArmandoCCCT08}, the TLS protocol and a basic version of the Kerberos protocol.  All experiments were performed on a server with 96 Intel(R) Xeon(R) Gold 6252 physical cores at 2.10GHz (192 logical cores), and 196GB of RAM. We show some details here for the simplified versions of Google SSO and Kerberos protocols; further details are available in the appendix. 
For each experiment, we specified the number of sessions and the depth of the search, and the \textit{par-depth} for the clustering of parallel search.
The protocols used for these experiments were all specified in the AnB format. These experiments were restricted to a \textit{par-depth} of 3, which seems to strike a reasonable balance between performance and memory consumption. 
For a protocol with $n$ steps per session, the length of the run of $k$ concurrent sessions of the protocol is bounded by $n*k$, which corresponds to the depth of search. 
Therefore, to prove the security of a protocol with $k$ steps for $n$ sessions, the depth of the search must be at least $n*k.$ 

\begin{figure}[t]
\includegraphics[width=0.5\linewidth]{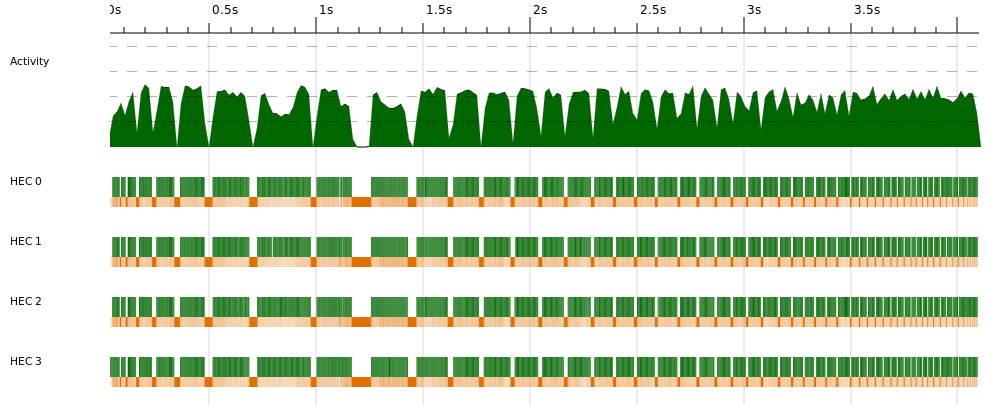}
\includegraphics[width=0.5\linewidth]{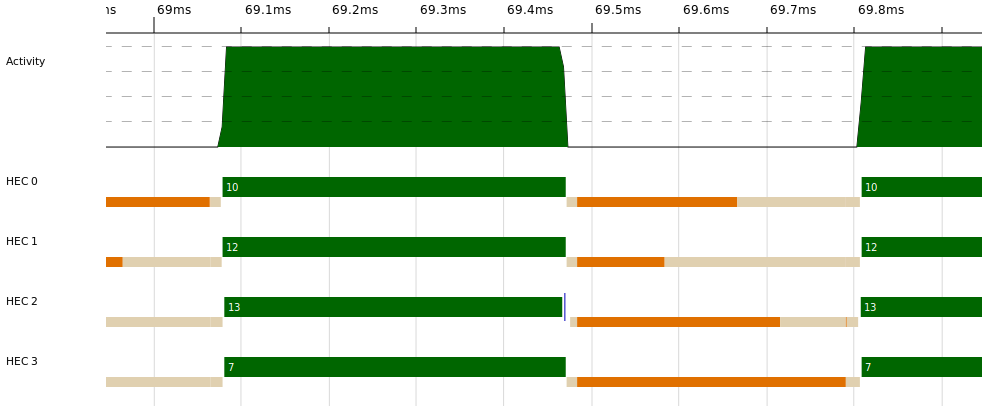}
\caption{Profiling workload distributions: (a) overall and (b) an individual activity.}
\label{fig:profile1}
\end{figure}

To check whether our parallelisation strategy did indeed distribute the work to multiple cores, we perform a profiling of runtime workload distributions among different cores. For this test, we used the Google Single Sign-On (SSO) protocol formalisation which comes with the OFMC distribution. The runtime profiles (Figure~\ref{fig:profile1}) were generated using the \lstinline[basicstyle=\small\ttfamily]{threadscope} software \cite{threadscope}. 
Figure~\ref{fig:profile1}a shows the overall profile. The green `bar' on the top is the overall workload of all cores combined, and the bars below (labelled HEC 0 - HEC 3) correspond to the activities for each core. The green bars represent the actual workload of the program being run. 
The orange bars denote time spent on garbage collection and the light orange bars represent idle time. As we can see in the figure, the work is distributed evenly across all cores, but there are gaps between activities.  
Figure~\ref{fig:profile1}b shows an individual activity in more detail. 
The workload component is only slightly longer than the combined GC and idle time. As we will see later, this ratio of workload vs overhead (GC + idle time) is consistently observed throughout our benchmarks. 

For the Kerberos protocol, we witnessed a huge increase in the number of states to verify. For 3 sessions and a search depth of 10, using a naive (unbufferred) strategy (e.g. \lstinline[basicstyle=\small\ttfamily]{parTree}), the verifier occasionally entered a state where it consumed a huge amount of memory (close to 190GB) and the program was terminated by the operating system. The problem became worse when moving to 4 sessions (with a search depth of 12); the memory consumption rose to around 130GB within 3 minutes of runtime. This is thus an example that shows the advantage of the buffered search tree strategy in PFMC. 

For this case study, we performed two sets of experiments. The first experiment used 3 sessions of the protocol, with a search depth of 10, while the second one increased the search depth to 18 (hence it covered all possible interleavings of actions from the 3 sessions). The purpose of this was primarily to see how the verification effort scales up with the increase of search depth. 

\begin{figure}[t]
        \includegraphics[width=0.5\linewidth]{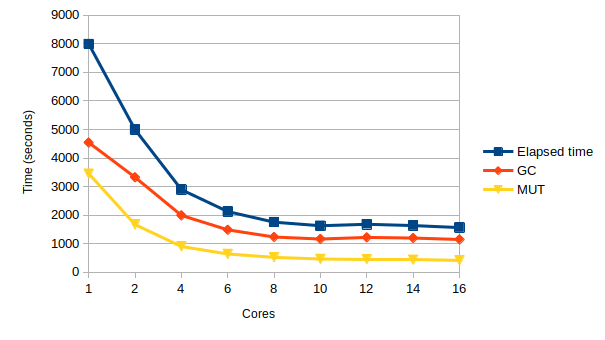}
        \includegraphics[width=0.5\linewidth]{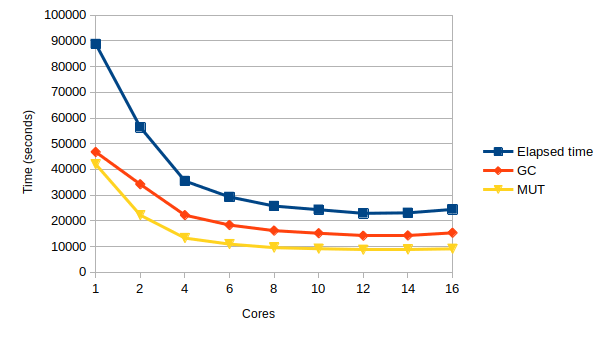}
        \vspace{-2mm}
    \caption{Execution time for the Kerberos protocol for 3 sessions with (a) a search depth of 10 and (b) a search depth of 18.}
    \label{fig:kerberos1-chart}
\end{figure}

The results of the experiments are summarised in Tables~\ref{tbl:kerberos} and \ref{tbl:kerberos2} and Figure~\ref{fig:kerberos1-chart}. The \textit{Total (elapsed)} column shows the total elapsed time (wall time).  
The \textit{Total (CPU)} column shows the total CPU time spent by all cores. The \textit{GC} column shows the (elapsed) time spent on garbage collection, and \textit{MUT} shows the actual productive time spent on the workload. The last column shows the maximum memory residency, i.e., the largest amount of memory used at any time.
The two experiments look remarkably similar. We see a steep improvement in elapsed time up to 10-12 cores, before the curves flatten out. 
The performance speed-ups in the best cases were 5.1 (for the first experiment, with 16 cores) and 3.8 (for the second experiment, with 14 cores). In the second experiment, the single-threaded OFMC took slightly over 24 hours, but the parallel version, in the best case, terminated after 6 hours or so.

\begin{table}[t]\scalebox{0.8}{
\begin{tabular}{r|r|r|r|r|r}
    Core\# & Total (elapsed) (s) & Total (CPU) (s) & GC  (s) & MUT (s) & Mem. res. (MB)\\
\hline
1 & 7990.432 & 7989.687 & 4539.233 & 3451.164 & 255.4\\
2 & 4992.650 & 9967.061 & 3324.002 & 1668.644 & 3609.8\\
4 & 2891.620 & 11422.235 & 1989.502 & 902.115 & 3954.3\\
6 & 2123.020 & 12443.051 & 1483.549 & 639.460 & 4122.6\\
8 & 1753.690 & 13493.416 & 1232.658 & 521.021 & 3930.3\\
10 & 1627.630 & 15456.142 & 1162.774 & 464.846 & 3861.1\\
12 & 1679.090 & 18870.687 & 1219.576 & 459.509 & 3829.5\\
14 & 1634.090 & 21193.560 & 1197.041 & 437.042 & 4120.1\\
16 & 1563.200 & 22944.665 & 1147.829 & 415.366 & 4204.5\\
\end{tabular}}
\caption{Kerberos protocol verification for 3 sessions and search depth 10.}
\label{tbl:kerberos}
\end{table}

\begin{table}[t]\scalebox{0.8}{
    \begin{tabular}{r|r|r|r|r|r}
        Core\# & Elapse time (s) & CPU time (s) & GC time (s) & MUT (s) & Mem. res. (MB)\\
    \hline
    1 & 88804.862  & 88796.110 & 46791.364 & 42013.268 & 2195.3\\
    2 & 56400.770 & 111668.513 & 34207.638 & 22193.131 & 6589.9 \\
    4 & 35479.590 & 133237.707 & 22194.304 & 13285.284 & 6766.1\\
    6 & 29285.870 & 157276.399 & 18348.008 & 10937.858 & 6689.4\\
    8 & 25753.490 & 177285.007 & 16184.376 & 9569.108 & 6742.4\\
    10 & 24328.190 & 201795.566 & 15171.428 & 9156.752 & 6795.9\\
    12 & 22916.410 & 221311.255 & 14186.897 & 8729.507 & 6823.9\\
    14 & 23114.540 & 254643.338 & 14284.919 & 8829.614 & 6864.3\\
    16 & 24443.140 & 304698.207 & 15330.385 & 9112.746 & 6868.6\\
    \end{tabular}}
    \caption{Kerberos protocol verification for 3 sessions and search depth 18.}
\label{tbl:kerberos2}
\end{table}

The long tails of distributions of performance gain in the previous experiments do not necessarily mean that we have reached the limit of parallelisation. Rather, it seems to be an effect on the limit of the \textit{par-depth} (and hence also the memory ceiling). Tweaking the \textit{par-depth} parameter slightly may result in better or worse performance. 
For the Kerberos verification, for 3 sessions and a search depth of 18, increasing the \textit{par-depth} to 4 from 3 gained us some improvement: the elapsed time was 21765.3 seconds (so about 1.12 speed up over the execution time using a \textit{par-depth} of 3), GC time 14271.5 seconds and MUT time 7494.8 seconds. However, the resident memory rose to 20GB (from 6.8GB). Raising the memory ceiling also allows more cores to be used to gain further performance improvement. For example, for the same Kerberos benchmark, with a \textit{par-depth} of 4, raising the number of cores to 20 resulted in a total elapsed time of 20789.7 seconds, with resident memory of 20.8GB.  
Generally, increasing the \textit{par-depth} results in higher memory consumption, but may allow all cores to be  maximally used at all times. When tested with a \textit{par-depth} of 6, the verification of Kerberos (3 sessions, search depth 18) exhausted the server's memory (196 GB) and was terminated by the operating system. 
These suggest that there are still performance improvements to be gained from \lstinline[basicstyle=\small\ttfamily]{parTreeBuffer} if we can reduce the overall memory footprint of PFMC, allowing us to use a greater \textit{par-depth}. However, generally, \lstinline[basicstyle=\small\ttfamily]{parTreeBuffer} is rather unsatisfactory as the memory consumption can grow unpredictably and crash the verifier. In our next strategies, we attempt to address this issue. 

Garbage collection currently seems to be the largest source of inefficiency in our experiments, taking up almost half of the execution time per core. This is, however, not due to the parallelisation, as it is also observed in the runtime for the single-core cases. Our conjecture is that this is an inherent issue with the search procedures underlying OFMC, which may involve creation of search nodes that end up not being evaluated and later discarded by the garbage collector.

\subsection{Enhanced parTreeBuffer}
This strategy was a modification to the original \lstinline{parTreeBuffer} strategy which led to a dramatic improvement in both speed and memory consumption. 
 The modification consists of two parts. First, the \textit{par-depth} limitation is removed, and \lstinline[basicstyle=\small\ttfamily]{parBuffer} is called at each node's children. This increases speed-up at the cost of memory overhead. However, this memory cost increase is offset by the second modification. The second change, which is the key modification, involves eagerly evaluating the {\em spine} of the list of children at each node, without evaluating each individual node in the list, prior to calling \lstinline[basicstyle=\small\ttfamily]{parBuffer}, as shown in Fig.~\ref{fig:alg3}.

\begin{figure}[t]
	\includegraphics[width=0.5\linewidth]{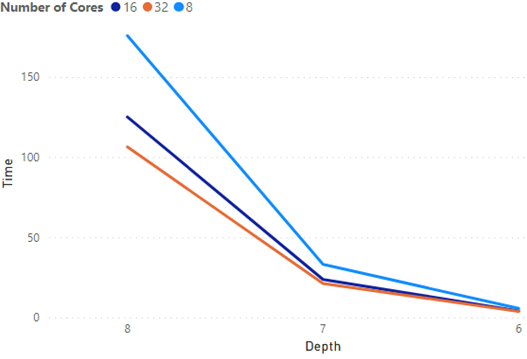}
	\includegraphics[width=0.5\linewidth]{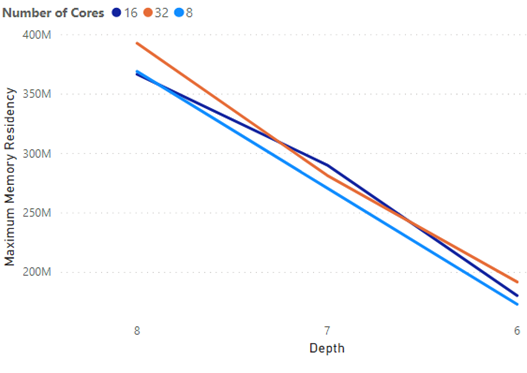}
	\caption{The effect of the depth and number of cores of the Kerberos protocol on (a) execution time and (b) memory residency, using the enhanced \lstinline[basicstyle=\small\ttfamily]{parTreeBuffer} strategy. }
	\label{fig:parBufferFigs}
\end{figure}

To understand why the second modification is significant, recall that due to the lazy evaluation of Haskell, when a function that returns a list is called, Haskell will stop evaluating the function as soon as the topmost constructor is evaluated, i.e., when the result is of the form \texttt{(x:l)}, where the head \texttt{x} and the tail \texttt{l} of the list are unevaluated expressions. In the case of PFMC, this list contains the successors of the underlying transition system encoding the protocol and the attacker moves. Deducing possible transitions may involve solving deducibility constraints, and as the number of sessions and the depth search grow, the accumulated constraints can be significant; we conjecture that evaluating this eagerly allows some of the large lazy terms to be simplified in advance, at little computation cost, which reduces the memory footprint significantly. 
This small change decreases the memory consumption of the strategy tenfold in many cases. 
Figure \ref{fig:parBufferFigs} shows the effect of the depth and number of cores on execution time and memory residency, using the basic Kerberos protocol.

\begin{figure}[t]
\begin{lstlisting}
parTreeBuffer :: Int -> Strategy a -> Strategy (Tree a)
parTreeBuffer _ strat (Node x []) = do 
	y <- strat x 
	return (Node y [])
parTreeBuffer c strat (Node x l) = do 
	y <- strat x 
	n <- rseq (length l) 
	l' <- parBuffer c (parTreeBuffer c strat) l
	return (Node y l')
\end{lstlisting} 
\vspace{-3mm}
\caption{parBuffer at each level, with eager evaluation of subchildren length.}
\label{fig:alg3}
\end{figure}

\subsection{parTreeChunkSubtrees}
This strategy aims at controlling exactly the overall number of sparks created. It uses a concept of {\em fuel} and fuel splitting, motivated by \cite{Totoo16}. 
It starts at the root node of the search tree with a given number of sparks (i.e., the fuel) to be created. It then divides this number between each subtree, evaluating nodes sequentially. 
When the number of sparks reaches one, a spark is created to evaluate the remaining subtrees in parallel. This method is extremely memory efficient and performs well on smaller trees. However, as the number of sparks requested increases, the sequential portion towards the root of the tree becomes larger. Therefore, this method is not suitable for large problem sizes. 
Nevertheless, the overall memory consumption of this strategy is extremely low, so it has potential applications where memory use is a priority, as well as for smaller problem sizes. An extension to this strategy which resolves this issue is \lstinline[basicstyle=\small\ttfamily]{hybridSubtrees}.

\subsection{hybridSubtrees}
\begin{figure}[t]
	\begin{center}
		\includegraphics[width=0.5\linewidth]{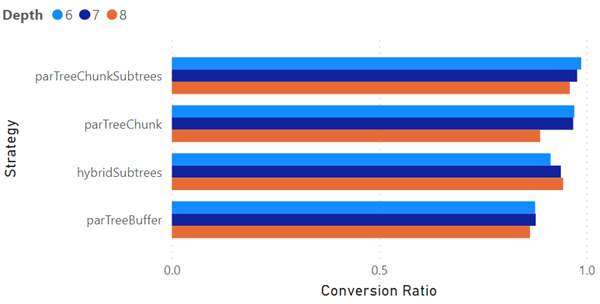}
	\end{center}
	\vspace{-5mm}
	\caption{Conversion Ratio using the \lstinline[basicstyle=\small\ttfamily]{hybridSubtrees} strategy for the Kerberos protocol.}
	\label{fig:conversionRatio}
\end{figure}

The \lstinline[basicstyle=\small\ttfamily]{hybridSubtrees} strategy
divides a certain number of sparks over the tree. However, instead of evaluating sequentially until it runs out of fuel, it recursively calls \lstinline[basicstyle=\small\ttfamily]{parBuffer} at each level, and divides the remaining sparks between its children. When the strategy can no longer create a full buffer, it creates a spark for the remaining subtree.
This strategy therefore has a strict upper bound on the degree of parallelism. It also affords better control over the trade-off between memory and performance in general. When given enough sparks, it behaves similarly to \lstinline[basicstyle=\small\ttfamily]{parTreeBuffer}. Given fewer sparks, the average granularity increases, the performance slows, and the memory consumption decreases. A benefit of sparking subtrees may also be that nodes deeper in the tree are less likely to require evaluation, meaning less work for the relatively overloaded subtree sparks. This is evidenced by its conversion rate when compared to other strategies, which indicates fewer fizzled and garbage collected sparks, as shown in Fig.~\ref{fig:conversionRatio}.

\begin{figure}[t]
	\includegraphics[width=0.5\linewidth]{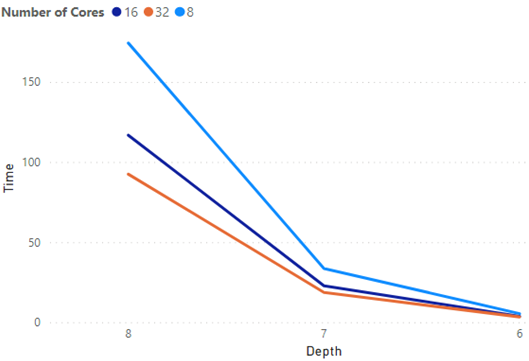}
	\includegraphics[width=0.5\linewidth]{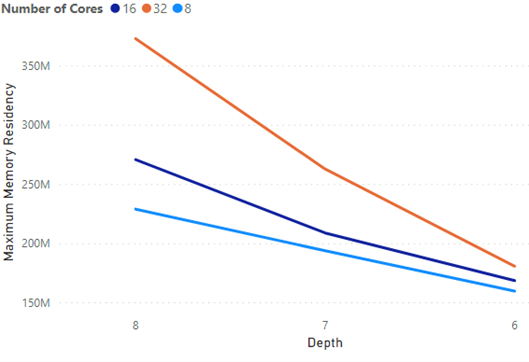}
	\caption{The effect of the depth and number of cores of the Kerberos protocol on (a) execution time and (b) memory residency, using the \lstinline[basicstyle=\small\ttfamily]{hybridSubtrees} strategy. }
	\label{fig:hybridSubtreesFigs}
\end{figure}

A limitation of this strategy is that the sparks are not evenly divided between subtrees, and the number of sparks created 
is usually significantly lower than the number requested. 
Overall, the strategy performed slightly better than \lstinline{parTreeBuffer} at certain problem sizes. However when testing very large problem sizes (4 sessions or depths greater than 10), spark creation as a proportion of the spark cap provided decreased, and it became difficult to assign adequate sparks to offer comparable speeds to \lstinline[basicstyle=\small\ttfamily]{parTreeBuffer} without thoroughly testing for a given depth to explore its sparking characteristics. Fig.~\ref{fig:hybridSubtreesFigs} shows the execution time and memory residency using this strategy on the Kerberos protocol.

\subsection{Strategies with Annotation}
The above strategies make assumptions about the shape of the tree which could impact performance for particular problems. For example, sparks could be left unused by \lstinline[basicstyle=\small\ttfamily]{hybridSubTrees} due to the unbalanced nature of the tree. Learning more about the search tree before applying a strategy could help to offset this. 
We experimented with several strategies using annotation methods inspired by \cite{Totoo16}, by annotating each node in the search tree with information about the number of subnodes, in order to achieve better sparking and load balancing characteristics. The core of these strategies was at each node developing a list of slightly different strategies to apply to each subtree, based on the annotation information. For example, in the case of fuel-splitting strategies such as \lstinline[basicstyle=\small\ttfamily]{parTreeChunkSubtrees} or \lstinline[basicstyle=\small\ttfamily]{hybridSubtrees}, annotation was used to ensure excess fuel was not passed to subtrees without sufficient nodes to make use of it.

However, in all cases these strategies did not justify their additional cost in performance and memory. This may be due to the fact that many nodes in the tree are not evaluated under normal conditions, but the annotation runs force a degree of evaluation in order to count nodes, and this overhead is not offset by any performance gains. It is also possible that the chosen method of annotation is inefficient, and better methods using, for example, a heuristic technique may perform better.

\subsection{Comparison}
The enhanced \lstinline[basicstyle=\small\ttfamily]{parTreeBuffer} strategy offers reliably good results for most problem sizes, with improved speed-up and memory consumption in all cases over the \lstinline[basicstyle=\small\ttfamily]{parTreeBuffer} strategy. For small trees, the simpler \lstinline[basicstyle=\small\ttfamily]{parTreeChunkSubtrees} performs better in both measures than any alternatives, but at these problem sizes the difference is not significant. The \lstinline[basicstyle=\small\ttfamily]{hybridSubTrees} strategy performs well if assigned adequate fuel, but this can become difficult for greater depths. 

\begin{figure}[t]
	\includegraphics[width=0.5\linewidth]{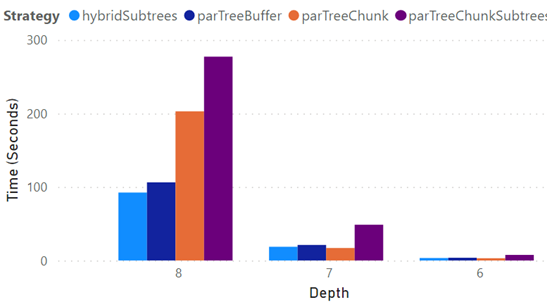}
	\includegraphics[width=0.5\linewidth]{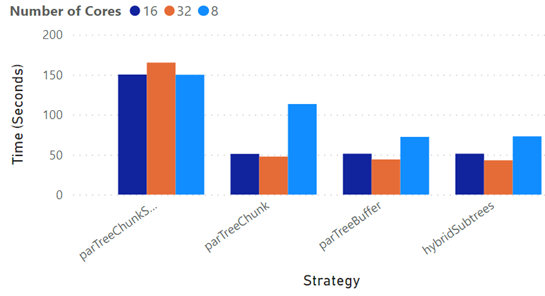}
	\caption{(a) Best execution time vs depth and (b) average execution time for varying numbers of cores for each of the strategies.}
	\label{fig:comparisons}
\end{figure}

\begin{figure}[t]
	\begin{center}
		\includegraphics[width=0.5\linewidth]{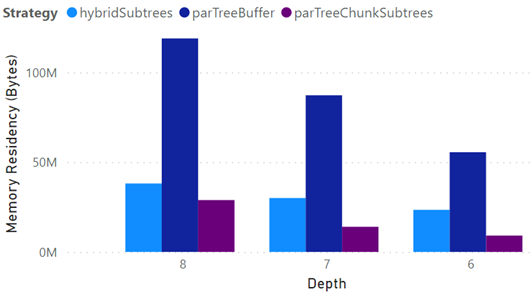}
	\end{center}
	\vspace{-7mm}
	\caption{Maximum memory residency vs depth for some of the strategies.}
	\label{fig:memory}
\end{figure}

For very large problems, where \lstinline[basicstyle=\small\ttfamily]{parTreeBuffer} may overflow memory due to excessive spark creation, \lstinline[basicstyle=\small\ttfamily]{hybridSubtrees} can also be used to limit parallelism while still benefiting from parallel speed-up. It performed consistently better in terms of memory consumption compared to \lstinline[basicstyle=\small\ttfamily]{parTreeBuffer} and offers more explicit control over the degree of parallelism. Figs.~\ref{fig:comparisons} and \ref{fig:memory} show the comparisons of execution time and memory for the different strategies.

These observations apply only to PFMC, and testing on a variety of other problems would be required for commenting on the more general performance characteristics of the strategies. Finer granularity of problem size at each node would make chunking strategies more applicable, whereas problems which require the evaluation of the entire tree at each execution would not benefit as much from the buffering approach and may be better served by fuel-splitting or some other strategy.

\subsection{Enabling the verification of protocols with algebraic operators}

OFMC also supports extensions of Dolev-Yao attacker models with algebraic operators. Some of these operators, such as XOR, are supported by default, but the user has the ability to add their own equational theory for custom operators. The custom theories, however, are supported only when the input is specified in the AVISPA Intermediate Format (IF) format~\cite{ModersheimV09}, which is not very user-friendly. OFMC does provide a translator from AnB format to IF format, but currently there are some issues in recognising the custom theories. We have made some modifications to the translator to allow us to experiment with verification of protocols with algebraic properties. In terms of performance speed-up, the results are in line with what we have seen in the case of Dolev-Yao attackers. Further details of these experiments are available in Appendix~\ref{sec:alg}. 
\section{Conclusion and Future Work}
\label{sec:conc}

Our preliminary results show that there is a significant improvement in moving towards moderate parallelisation of security protocol verification based on symbolic constraint solving. We managed to achieve 3-5 times speed up compared to the sequential verifier, utilising between 4-16 cores, on benchmarks with 3 sessions. Beyond 16 cores the performance improvement does not seem substantial. This is partly due to our algorithm limiting the memory usage, preventing more tasks to be executed in parallel, but in some cases it could also be that the problem has hit a limit of parallelisation, e.g., some parts of the search could not be parallelised due to inherence dependencies.

One major issue in scaling up the protocol verification is controlling the memory consumption, while attaining a good degree of parallelisation. To this end we have been experimenting with various buffering and chunking strategies. The extended \lstinline[basicstyle=\small\ttfamily]{parTreeBuffer} and the \lstinline[basicstyle=\small\ttfamily]{hybridSubTree} strategies (Section~\ref{sec:strategies}) seem to so far provide a good balance of memory consumption and performance, with \lstinline[basicstyle=\small\ttfamily]{parTreeBuffer} generally performing better, but with a worse memory footprint. 

For future work, we plan to explore annotation methods to achieve better load balancing in sparking the search trees, perhaps by moving to a limited shared-memory concurrent setting. We also plan to apply our light-weight parallelisation approach to improve the performance of other protocol verifiers, e.g., Tamarin.

\bibliographystyle{abbrv}
\bibliography{biblio}

\begin{thebibliography}{10}

\bibitem{AbadiG99}
M.~Abadi and A.~D. Gordon.
\newblock A calculus for cryptographic protocols: The spi calculus.
\newblock {\em Inf. Comput.}, 148(1):1--70, 1999.

\bibitem{AlmousaMV15}
O.~Almousa, S.~M{\"{o}}dersheim, and L.~Vigan{\`{o}}.
\newblock Alice and bob: Reconciling formal models and implementation.
\newblock In {\em Programming Languages with Applications to Biology and
  Security - Essays Dedicated to Pierpaolo Degano on the Occasion of His 65th
  Birthday}, volume 9465 of {\em {LNCS}}, pages 66--85. Springer, 2015.

\bibitem{ArapinisDK08}
M.~Arapinis, S.~Delaune, and S.~Kremer.
\newblock From one session to many: Dynamic tags for security protocols.
\newblock In {\em Logic for Programming, Artificial Intelligence, and
  Reasoning, 15th International Conference, {LPAR} 2008. Proceedings}, volume
  5330 of {\em {LNCS}}, pages 128--142. Springer, 2008.

\bibitem{ArmandoCC16}
A.~Armando, R.~Carbone, and L.~Compagna.
\newblock {SATMC:} a sat-based model checker for security protocols, business
  processes, and security apis.
\newblock {\em Int. J. Softw. Tools Technol. Transf.}, 18(2):187--204, 2016.

\bibitem{ArmandoCCCT08}
A.~Armando, R.~Carbone, L.~Compagna, J.~Cu{\'{e}}llar, and L.~Tobarra.
\newblock Formal analysis of {SAML} 2.0 web browser single sign-on: breaking
  the saml-based single sign-on for google apps.
\newblock In {\em Proceedings of the 6th {ACM} Workshop on Formal Methods in
  Security Engineering, {FMSE} 2008.}, pages 1--10. {ACM}, 2008.

\bibitem{BarbosaBBBCLP19}
M.~Barbosa, G.~Barthe, K.~Bhargavan, B.~Blanchet, C.~Cremers, K.~Liao, and
  B.~Parno.
\newblock Sok: Computer-aided cryptography.
\newblock {\em {IACR} Cryptol. ePrint Arch.}, 2019:1393, 2019.

\bibitem{BasinMV05}
D.~A. Basin, S.~M{\"{o}}dersheim, and L.~Vigan{\`{o}}.
\newblock Algebraic intruder deductions.
\newblock In {\em Logic for Programming, Artificial Intelligence, and
  Reasoning, 12th International Conference, {LPAR} 2005, Proceedings}, volume
  3835 of {\em {LNCS}}, pages 549--564. Springer, 2005.

\bibitem{Blanchet11}
B.~Blanchet.
\newblock Using horn clauses for analyzing security protocols.
\newblock In {\em Formal Models and Techniques for Analyzing Security
  Protocols}, volume~5 of {\em Cryptology and Information Security Series},
  pages 86--111. {IOS} Press, 2011.

\bibitem{Blanchet16}
B.~Blanchet.
\newblock Modeling and verifying security protocols with the applied pi
  calculus and proverif.
\newblock {\em Found. Trends Priv. Secur.}, 1(1-2):1--135, 2016.

\bibitem{CervesatoDLMS99}
I.~Cervesato, N.~A. Durgin, P.~Lincoln, J.~C. Mitchell, and A.~Scedrov.
\newblock A meta-notation for protocol analysis.
\newblock In {\em Proceedings of the 12th {IEEE} Computer Security Foundations
  Workshop, {CSFW} 1999}, pages 55--69. {IEEE} Computer Society, 1999.

\bibitem{Cheval14}
V.~Cheval.
\newblock {APTE:} an algorithm for proving trace equivalence.
\newblock In {\em Tools and Algorithms for the Construction and Analysis of
  Systems - 20th International Conference, {TACAS} 2014. Proceedings}, volume
  8413 of {\em {LNCS}}, pages 587--592. Springer, 2014.

\bibitem{Cheval18SP}
V.~Cheval, S.~Kremer, and I.~Rakotonirina.
\newblock {DEEPSEC:} deciding equivalence properties in security protocols
  theory and practice.
\newblock In {\em 2018 {IEEE} Symposium on Security and Privacy, {SP} 2018,
  Proceedings}, pages 529--546. {IEEE} Computer Society, 2018.

\bibitem{Cortier18ESORICS}
V.~Cortier, A.~Dallon, and S.~Delaune.
\newblock Efficiently deciding equivalence for standard primitives and phases.
\newblock In {\em Computer Security - 23rd European Symposium on Research in
  Computer Security, {ESORICS} 2018, Proceedings, Part {I}}, volume 11098 of
  {\em {LNCS}}, pages 491--511. Springer, 2018.

\bibitem{CortierDL06}
V.~Cortier, S.~Delaune, and P.~Lafourcade.
\newblock A survey of algebraic properties used in cryptographic protocols.
\newblock {\em J. Comput. Secur.}, 14(1):1--43, 2006.

\bibitem{DolevY83}
D.~Dolev and A.~C. Yao.
\newblock On the security of public key protocols.
\newblock {\em {IEEE} Trans. Information Theory}, 29(2):198--207, 1983.

\bibitem{Durgin99}
N.~A. Durgin, P.~Lincoln, J.~C. Mitchell, and A.~Scedrov.
\newblock Undecidability of bounded security protocols.
\newblock In {\em {W}orkshop on {F}ormal {M}ethods and {S}ecurity {P}rotocols},
  1999.

\bibitem{Kanovich14}
M.~I. Kanovich, T.~B. Kirigin, V.~Nigam, and A.~Scedrov.
\newblock Bounded memory dolev-yao adversaries in collaborative systems.
\newblock {\em Inf. Comput.}, 238:233--261, 2014.

\bibitem{threadscope}
S.~Marlow, D.~Jones, and S.~Singh.
\newblock threadscope (software package).
\newblock \url{https://wiki.haskell.org/ThreadScope}.

\bibitem{MarlowNJ11}
S.~Marlow, R.~Newton, and S.~L.~P. Jones.
\newblock A monad for deterministic parallelism.
\newblock In {\em Proceedings of the 4th {ACM} {SIGPLAN} Symposium on Haskell,
  Haskell 2011, Tokyo, Japan, 22 September 2011}, pages 71--82. {ACM}, 2011.

\bibitem{MeierSCB13}
S.~Meier, B.~Schmidt, C.~Cremers, and D.~A. Basin.
\newblock The {TAMARIN} prover for the symbolic analysis of security protocols.
\newblock In {\em Computer Aided Verification - 25th International Conference,
  {CAV} 2013. Proceedings}, volume 8044 of {\em {LNCS}}, pages 696--701.
  Springer, 2013.

\bibitem{ModersheimV09}
S.~M{\"{o}}dersheim and L.~Vigan{\`{o}}.
\newblock The open-source fixed-point model checker for symbolic analysis of
  security protocols.
\newblock In {\em Foundations of Security Analysis and Design V, {FOSAD}
  2007/2008/2009 Tutorial Lectures}, volume 5705 of {\em {LNCS}}, pages
  166--194. Springer, 2009.

\bibitem{RusinowitchT01}
M.~Rusinowitch and M.~Turuani.
\newblock Protocol insecurity with finite number of sessions is np-complete.
\newblock In {\em 14th {IEEE} Computer Security Foundations Workshop {(CSFW-14}
  2001)}, pages 174--187. {IEEE} Computer Society, 2001.

\bibitem{ThayerHG99}
F.~J. Thayer, J.~C. Herzog, and J.~D. Guttman.
\newblock Strand spaces: Proving security protocols correct.
\newblock {\em J. Comput. Secur.}, 7(1):191--230, 1999.

\bibitem{TiuNH16}
A.~Tiu, N.~Nguyen, and R.~Horne.
\newblock {SPEC:} an equivalence checker for security protocols.
\newblock In {\em Programming Languages and Systems - 14th Asian Symposium,
  {APLAS} 2016, Proceedings}, volume 10017 of {\em {LNCS}}, pages 87--95, 2016.

\bibitem{Totoo16}
P.~Totoo.
\newblock {\em Parallel evaluation strategies for lazy data structures in
  {H}askell}.
\newblock PhD thesis, Heriot-Watt University, 2016.

\bibitem{DeursenR08}
T.~van Deursen and S.~Radomirovic.
\newblock Attacks on {RFID} protocols.
\newblock {\em {IACR} Cryptol. ePrint Arch.}, page 310, 2008.

\end{thebibliography}

\appendix 
\section{Case studies}

We show here more details of the tests we performed on some example protocols. 
These example protocols are from the original distribution of OFMC (and also included in the distribution of PFMC).  

\subsection{A simplified Kerberos protocol}

\begin{figure}
\begin{lstlisting}
Protocol: Basic_Kerberos  # Bounded-verified

Types: Agent C,a,g,s;
      Number N1,N2,T1,T2,T3,Payload,tag;
      Function hash,sk;
      Symmetric_key KCG,KCS

Knowledge: C: C,a,g,s,sk(C,a),hash,tag;
	   a: a,g,hash,C,sk(C,a),sk(a,g),tag;
	   g: a,g,sk(a,g),sk(g,s),hash,tag;
	   s: g,s,sk(g,s),hash,tag

Actions:

C -> a: C,g,N1
a -> C: {| KCG, C, T1 |}sk(a,g), 
       {| KCG, N1, T1, G |}sk(C,a)
C -> g: {| KCG, C, T1 |}sk(a,g), {|C,T1|}KCG, s,N2
g -> C: {| KCS, C, T2 |}sk(g,s), {| KCS, N2, T2, s |}KCG
C -> s: {| KCS, C, T2 |}sk(g,s), {| C, T3 |}KCS
s -> C: {|T3|}KCS, {|tag,Payload|}KCS

Goals:
s *->* C: Payload

\end{lstlisting}
\caption{A simplified version of the Kerberos protocol}
\label{fig:kerberos}
\end{figure}

Figure~\ref{fig:kerberos} shows the formalisation of a simplified version of the Kerberos protocol in the ``Alice and Bob'' (AnB) notation of OFMC. It abstracts away some aspects of the actual protocol, such as the authentication tags. The goal states that the protocol establishes a confidential and authentic channel between $s$ and $C$. 
We have provided details of the performance evaluation of Kerberos in Section~\ref{sec:strategies} so we will not repeat it here.

\subsection{Google Single Sign-On protocol}

\begin{figure}
\begin{lstlisting}
Protocol: SingleSignOn 
   # the flawed version of Google's SSO from before 2008 [Armando et al.]
   # in comments the standard specification (which is safe)
   
Types: Agent C,idp,SP;
      Number URI,ID,Data;
      Function h,sk
   
Knowledge: C: C,idp,SP,pk(idp);
          idp: C,idp,pk(idp),inv(pk(idp));
          SP: idp,SP,pk(idp)
          where SP!=C, SP!=idp, C!=idp
Actions:
   
   [C] *->* SP  : C,SP,URI
    SP *->* [C] : C,idp,SP,ID,URI
   
    C  *->* idp : C,idp,SP,ID,URI
   # google:
   idp *->* C   : {C,idp}inv(pk(idp)),URI
   # standard:
   #idp *->* C   : ID,SP,idp,{ID,C,idp,SP}inv(pk(idp)),URI
   
   # google:
   [C] *->* SP  : {C,idp}inv(pk(idp)),URI
   # standard:
   #[C] *->* SP  : ID,SP,idp,{ID,C,idp,SP}inv(pk(idp)),URI
   SP *->* [C] : Data,ID
   
Goals:
   
   SP authenticates C on URI
   C authenticates SP on Data
   Data secret between SP,C
\end{lstlisting}
\caption{A flawed version of Google Single Sign-On protocol. Source: OFMC distribution.}
\label{fig:sso}
\end{figure}

The attack can be found within 2 sessions. The (simplified) protocol is given in 
Figure~\ref{fig:sso}. The constructor \lstinline{pk} denotes the public key constructor; \lstinline{idp} is the identity provider, \lstinline{C} is the client, \lstinline{S} is the service provider and
the private key that corresponds to a public key \lstinline{pk(X)} is denoted by \lstinline{inv(pk(X))}. 
The flaw in Google's implementation of the protocol (which was based on SAML 2.0) is in step 4 (line 20 in Figure~\ref{fig:sso}): Google's implementation omitted certain information, such as the unique identifier of the authentication request (the \lstinline{ID} variable in the protocol) and the identity of the service provider \lstinline{SP}. We will not go into the details of the attack; the interested reader can consult~\cite{ArmandoCCCT08}.
This example shows a flawed Google Single Sign-On protocol (SSO); the flaw was discovered 
by Armando et. al.~\cite{ArmandoCCCT08}  using the SATMC model checker~\cite{ArmandoCC16}. 
There is an attack when two sessions of the protocol are running concurrently. Thus, to prove or disprove the security goals, we set the search depth to 12 (since each session can have at most 6 steps). For this and the next two case studies, we set the \textit{par-depth} to 3. 

\begin{figure}
\includegraphics[width=0.5\linewidth]{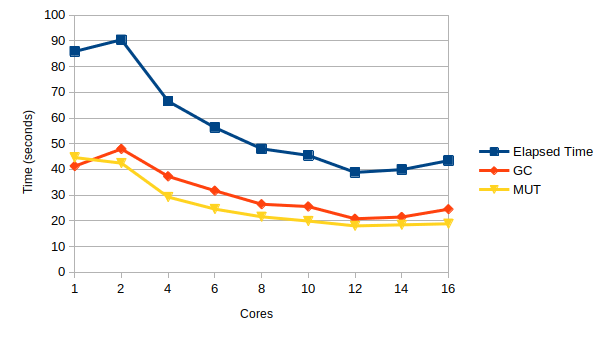}
\includegraphics[width=0.5\linewidth]{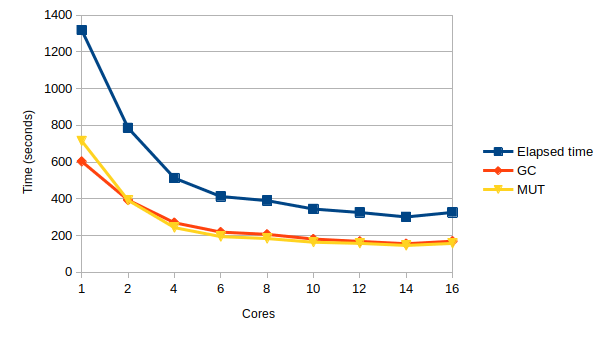}
\caption{Execution time for (a) the SSO protocol verification and (b) the TLS protocol.}
\label{fig:sso-chart}
\end{figure}

We repeated the experiment 8 times, increasing the number of cores by 2 with every iteration. The result is given in Table~\ref{tbl:sso}. 
The experiment results are given in Table~\ref{tbl:sso}. 
%The \textit{Elapsed time} column shows the total elapsed time (wall time) -- which is what matters from a user's perspective. The \textit{CPU time} column shows the total CPU time spent by all cores. The \textit{GC} column shows the (elapsed) time spent on garbage collection, and \textit{MUT} shows the actual productive time spent on the workload. The last column shows the maximum memory residency, i.e., the largest amount of memory used at any time. 
Figure~\ref{fig:sso-chart}a shows the trend in a chart. 

We note that there is a significant improvement from a single core to four cores. This trend continues to 12 cores, after which point the GC overhead seems to outweigh the performance gain. 
Interestingly, the GC overhead does not seem to become worse with the addition of cores; in fact, it seems to improve. However, GC accounts for almost half of the execution time. This may indicate that the sequential algorithm itself is quite inefficient. 
The performance gain through parallelism is offset by the significant increase of memory residency, so it would seem that availability of memory is a limiting factor in scaling up the verification. 
The fast growth in memory consumption is what originally motivated us to develop a buffered search strategy. As the table shows, the memory residency in this case stays relatively constant, despite the increase of the number of cores used.

\begin{table}\scalebox{0.9}{
\begin{tabular}{r|r|r|r|r|r}
Core\# & Elapsed time (s) & CPU time (s) & GC time (s) & MUT (s) & Mem. res. (MB)\\
\hline
1 & 85.909 & 85.902	& 41.283 & 44.625 & 4.3\\
2 & 90.450 & 180.114 & 47.981 & 42.462 & 108.2\\
4 & 66.530 & 254.309 & 37.285 & 29.239 & 122.8\\
6 & 56.280 & 304.819 & 31.705 & 24.568 & 146.3\\
8 & 48.040 & 328.983 & 26.460 & 21.573 & 131.8\\
10 & 45.480 & 374.103 & 25.568 & 19.907 & 138.0\\
12 & 38.840 & 364.404 & 20.804 & 18.027 & 143.0\\
14 & 39.930 & 425.004 & 21.485 & 18.439 & 142.6\\
16 & 43.440 & 524.929 & 24.537 & 18.854 & 147.4
\end{tabular}}
\caption{Results for the Google SSO verification for 2 sessions and search depth of 12.}
\label{tbl:sso}
\end{table}

\subsection{TLS protocol}

\begin{figure}

\begin{lstlisting}
Protocol: TLS  # Bounded-verified

Types: Agent A,B,s;
      Number NA,NB,Sid,PA,PB,PMS;
      Function pk,hash,clientK,serverK,prf
   
Knowledge: A: A,pk(A),pk(s),inv(pk(A)),{A,pk(A)}inv(pk(s)),B,hash,clientK,serverK,prf;
       B: B,pk(B),pk(s),inv(pk(B)),{B,pk(B)}inv(pk(s)),hash,clientK,serverK,prf
   
Actions:
   
   A->B: A,NA,Sid,PA
   B->A: NB,Sid,PB,
         {B,pk(B)}inv(pk(s))
   A->B: {A,pk(A)}inv(pk(s)),
       {PMS}pk(B),
       {hash(NB,B,PMS)}inv(pk(A)),
       {|hash(prf(PMS,NA,NB),A,B,NA,NB,Sid,PA,PB,PMS)|}
         clientK(NA,NB,prf(PMS,NA,NB))
   B->A: 	{|hash(prf(PMS,NA,NB),A,B,NA,NB,Sid,PA,PB,PMS)|}
         serverK(NA,NB,prf(PMS,NA,NB))
   
Goals:
   
   B authenticates A on prf(PMS,NA,NB)
   A authenticates B on prf(PMS,NA,NB)
   prf(PMS,NA,NB) secret between A,B
   
\end{lstlisting}   
\caption{TLS handshake protocol. Source: OFMC distribution.}
\label{fig:tls}
\end{figure}

For this case study, we verify the TLS handshake protocol, for 3 concurrent sessions. Figure~\ref{fig:tls} shows the formalisation of a simplified version of TLS in OFMC. Here we omit an explicit formalisation of certificates and certificate authorities. Digital signatures are also modelled using public key encryption, i.e., a digital signature is just an encryption using the private key. The various parameters in the TLS handshake protocol are also abstracted away as the (random) number \lstinline{PMS}. 
The function symbol \lstinline{hash} denotes a secure cryptographic hash function, the symbol \lstinline{clientK} denotes the function for constructing the (symmetric) encryption key for the client, and \lstinline{serverK} denotes the function for constructing the encryption key for the server. We do not  explicitly model the MAC keys.   

\begin{figure}
   \begin{center}
       \includegraphics[width=13cm]{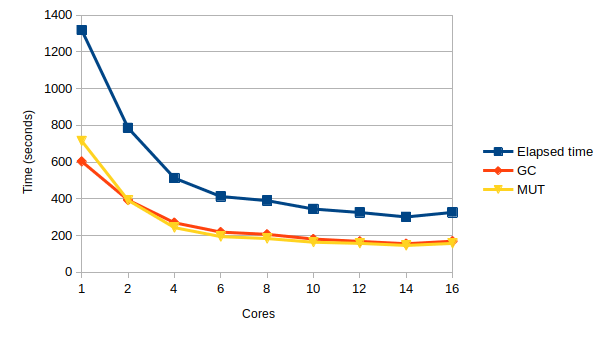}
   \end{center}
   \caption{Execution time TLS verification}
   \label{fig:tls-chart}
\end{figure}

The TLS handshake protocol has six steps, so to verify three sessions of the protocol, we need to consider a search depth of at least 18. As in the case with SSO, we tested PFMC on this protocol with increasing numbers of cores. The results are summarised in Table~\ref{tbl:tls} and Figure~\ref{fig:tls-chart}. 
Again we observe a similar pattern of a significant reduction in elapsed time up to around 10-12 cores, before the curve flattens. In this case however, we observe a steeper decline in total elapsed time, with around 4.4 times speed up when run on 14 cores. The GC and the MUT time are roughly the same throughout. 

\begin{table}\scalebox{0.9}{
\begin{tabular}{r|r|r|r|r|r}
    Core\# & Elapse time (s) & CPU time (s) & GC time (s) & MUT (s) & Mem. res. (MB)\\
\hline
1 & 1318.527 & 1318.397 & 603.617 & 714.908 & 19.8\\
2  & 786.020 & 1555.283 & 393.825 & 392.192 & 1203.3\\
4  & 512.580 & 1894.152 & 269.609 & 242.965 & 1412.7 \\
6  & 412.580 & 2153.811 & 218.471 & 194.106 & 1429.8\\
8  & 389.970 & 2555.735 & 206.417 & 183.541 & 1537.3\\
10 & 344.250 & 2692.709 & 180.713 & 163.527 & 1617.4\\
12 & 325.430 & 2925.672 & 168.043 & 157.379 & 1564.3\\
14 & 301.230 & 3085.562 & 155.248 & 145.974 & 1535.4\\
16 & 326.050 & 3722.634 & 168.967 & 157.044 & 1531.1\\
\end{tabular}}
\caption{Results for the TLS verification for 3 sessions and search depth of 12.}
\label{tbl:tls}
\vspace{-5mm}
\end{table}

\section{Enabling the Verification of Protocols with Algebraic Operators}
\label{sec:alg}

Most approaches for protocol verification usually assume the existence of a \textit{free algebra}.
 Free algebras do not include destructors. 
 The deduction capabilities of the intruder are handled by a set of deduction rules. However, certain algebraic operators 
used in cryptography cannot be modelled using only deduction rules. These operators require an
 \textit{equational theory}, which specifies a set of equations that describe the behaviour of the operator.
 An example of such an algebraic operator is the XOR (exclusive-Or) operator.
The equational theory of XOR does not result in a convergent rewriting system, due to the associativity 
and commutativity equations. 

OFMC has some limited support for handling algebraic operators. It allows users to specify a custom theory
 file which describes equational theories. The drawback is that support for the theory files is not
  complete. The theories for some standard algebraic operators are built into the code, such as XOR and
   encryption operators. However, using theory files to specify custom algebraic operators currently
   only works with models specified in the \textit{Intermediate Format
      (IF)}. There is an option for translating an AnB model into an IF model, but this still has some difficulties.
        
We extended PFMC to
provide support for users to use custom theory files using the existing OFMC translation to IF
  files. Using the existing translation from AnB to IF models is not straight-forward; there are several
   problems to overcome in order to use the translation for checking models with custom algebraic operators. 
   
In the new version of PFMC, users can specify the properties of several custom operators in a separate theory
    file and then use the translate option originally provided in OFMC to create an IF model where the custom operators are recognised.
Due to the current incomplete state of OFMC's custom theory option, several manual modifications are required in order to bypass errors thrown by OFMC when performing standard checks on the model which appear to fail when the custom theory files are not considered. For example, a common error thrown is that a secret X is never known by an agent A, although A may be able to deduce the secret using the equational theory properties.
           After translation, the resulting IF model
           must then be manually edited to remove the extra modifications. In most cases, this is far less
            tedious and error-prone than creating an entire IF model by hand, since IF models are significantly more
            complex than AnB models. 

The translation produced has a default number of sessions of two. Unlike AnB models, the number of
 sessions is fixed in IF models. In order to increase the number of sessions, separate IF files must be created for different numbers of sessions, by specifying the number required at the translation stage. The manual modifications are identical for all the files, so can simply be specified once and copied to the other files.

For several examples, the manual modifications are
 simple to perform, such as for the CH07 protocol from \cite{DeursenR08} and the Salary Sum and Shamir-Rivest-Adleman Three Pass protocols from \cite{CortierDL06}. Some models do not
  require any modifications, such as the LAK protocol described in \cite{DeursenR08}. The issue is usually in cases where the
   protocol contains information contained within custom operator functions, or inside nested operators,
    for which OFMC assumes the inner information cannot be extracted.
    
There are, however, certain models for which it is not straight-forward to translate them into IF models even with manual modifications.
 One such example is Bull's authentication protocol (described in \cite{CortierDL06}), for which it is not trivial to avoid the
  OFMC error checks.     
While in general it would be possible to prevent all the initial checks from being performed, a more robust solution for future work would be to modify the tool to use the custom theory files when translating.

\subsection{Experiments}
\begin{figure}
        \includegraphics[width=0.5\linewidth]{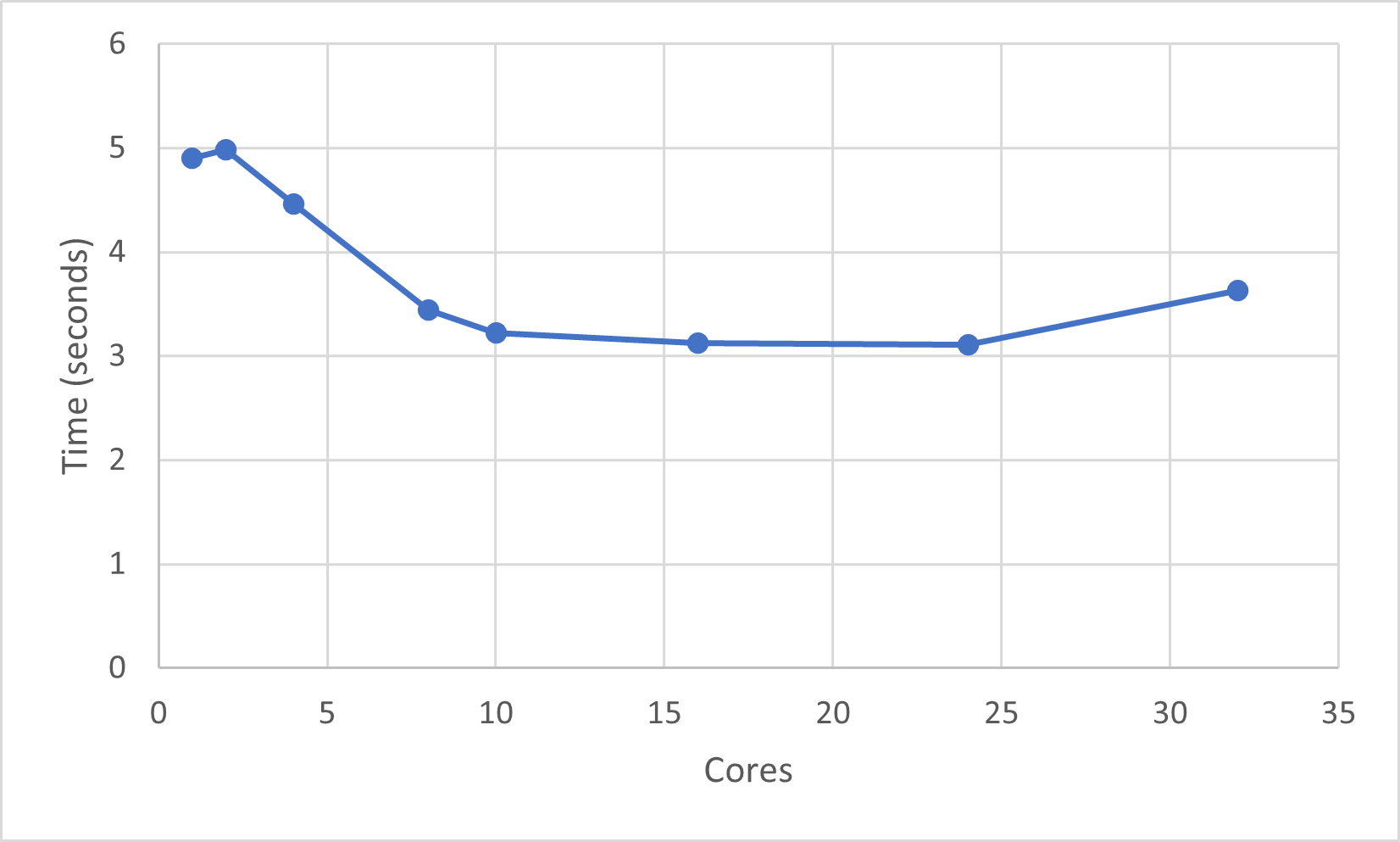}
        \includegraphics[width=0.5\linewidth]{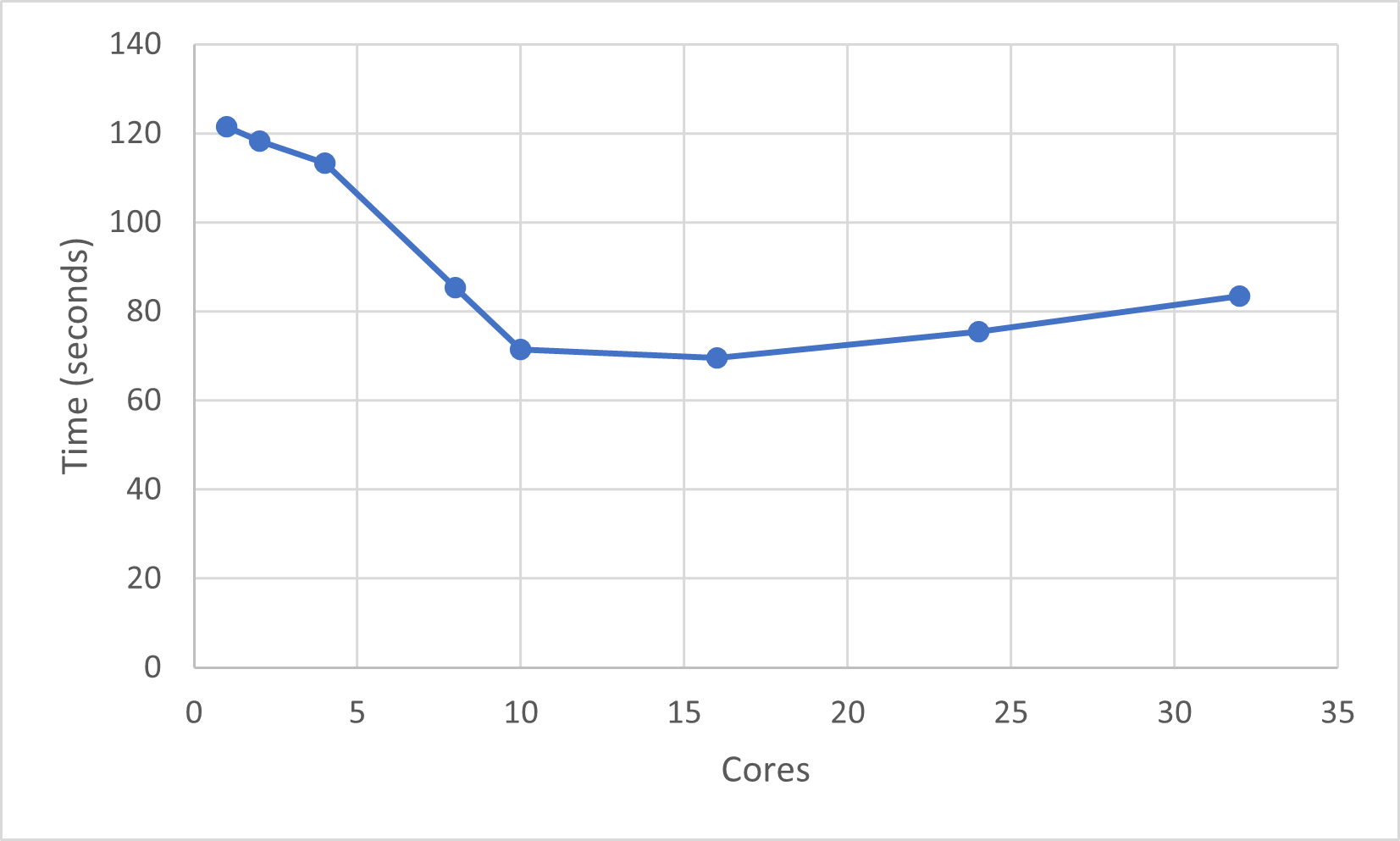}
    \caption{Execution time for the IKA protocol verification a search depth of 10 with (a) 3 sessions and (b) 4 sessions.}
    \label{fig:IKA1-chart}
\end{figure}

\begin{figure}
        \includegraphics[width=0.5\linewidth]{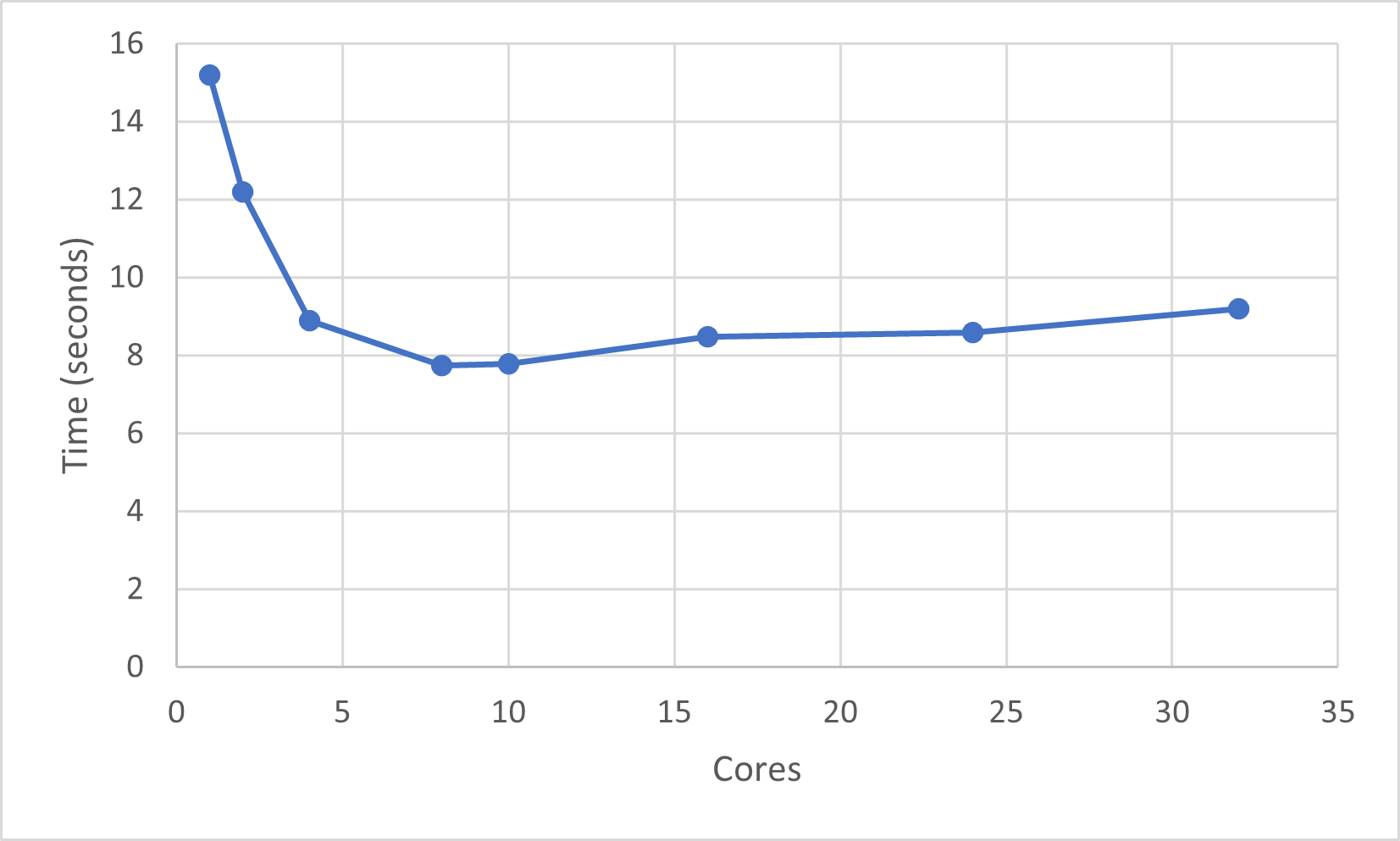}
        \includegraphics[width=0.5\linewidth]{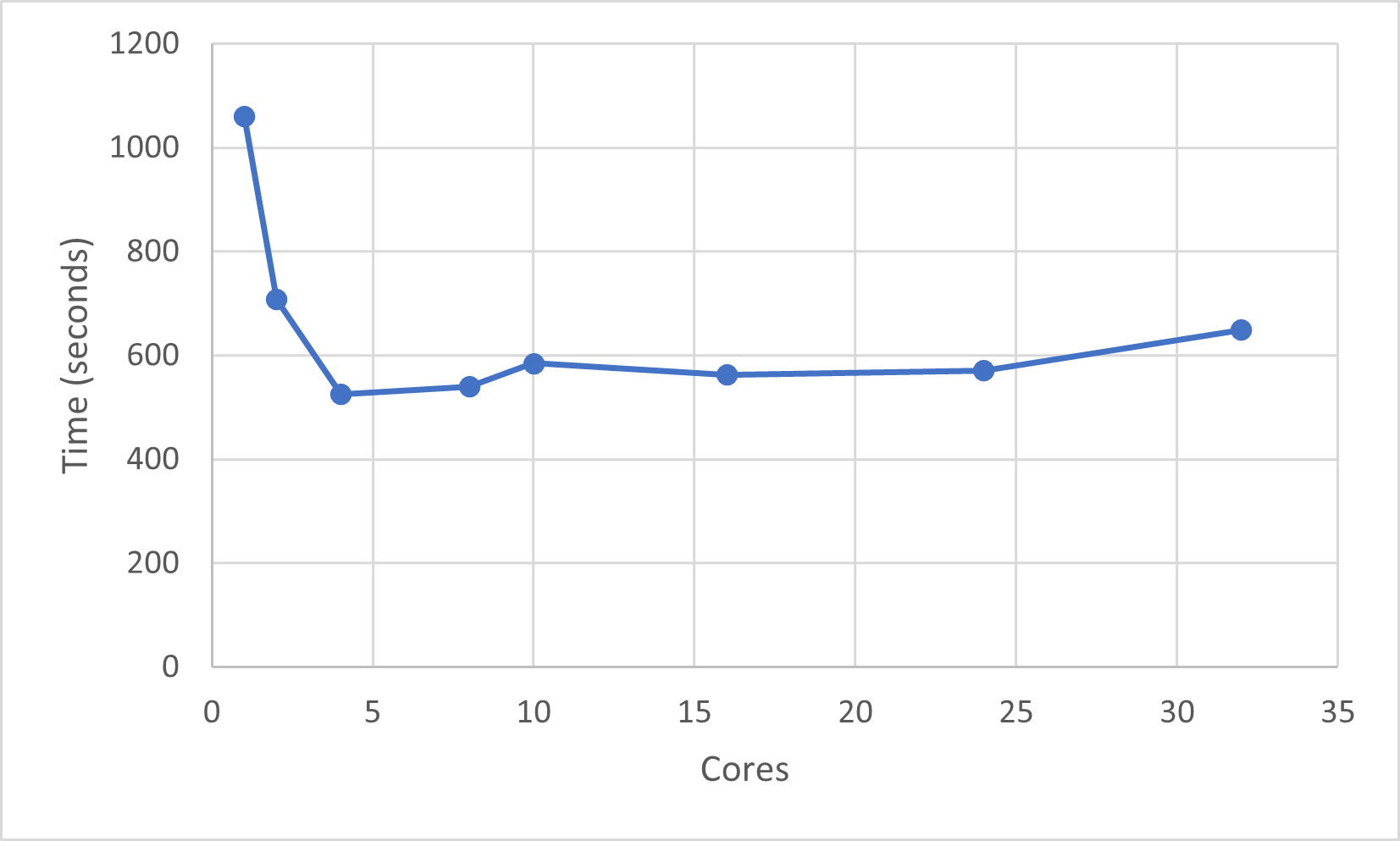}
    \caption{Execution time for the SRA Three Pass protocol verification a search depth of 10 with (a) 3 sessions and (b) 4 sessions.}
    \label{fig:SRA1-chart}
\end{figure}

\begin{figure}
        \includegraphics[width=0.5\linewidth]{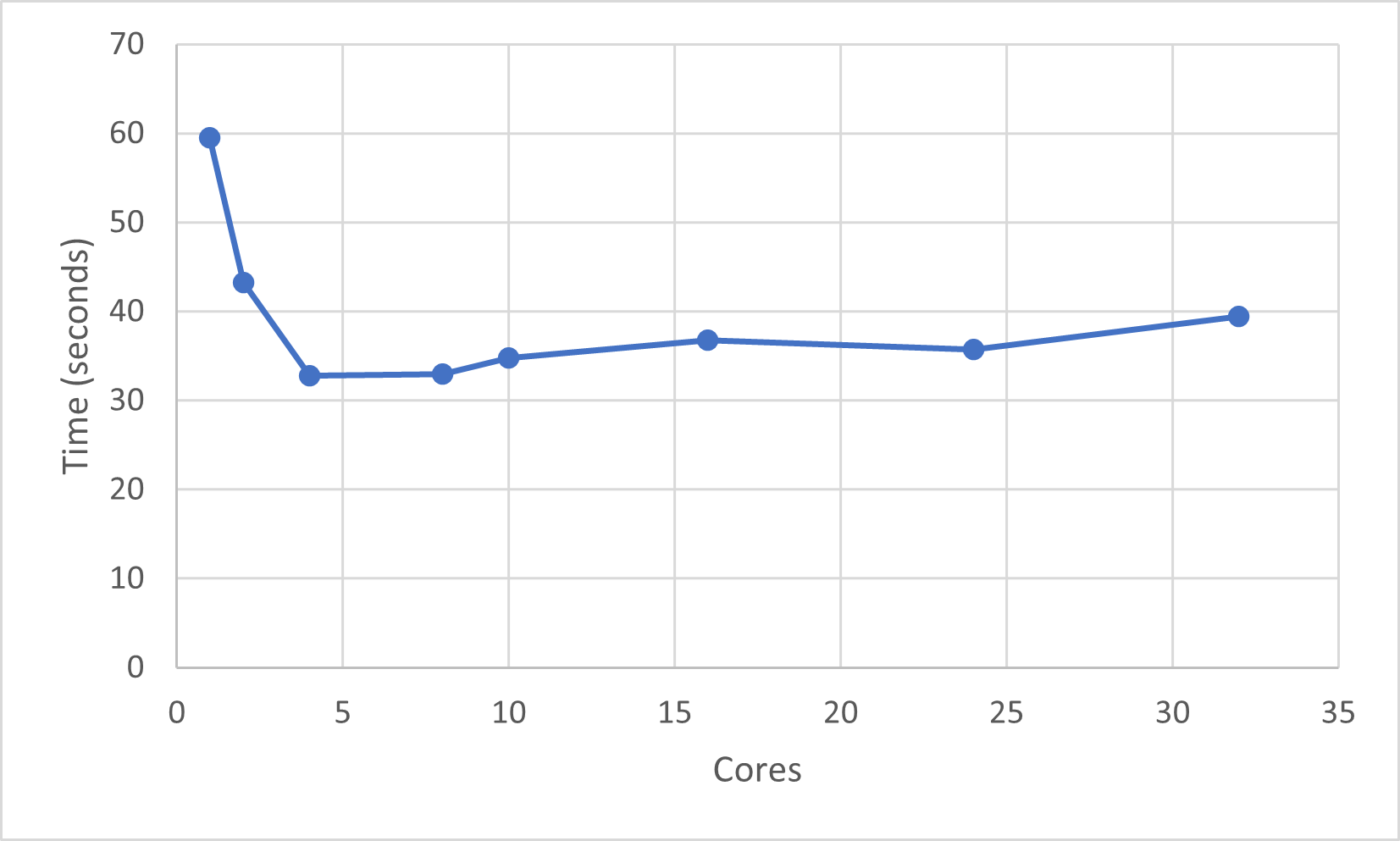}
        \includegraphics[width=0.5\linewidth]{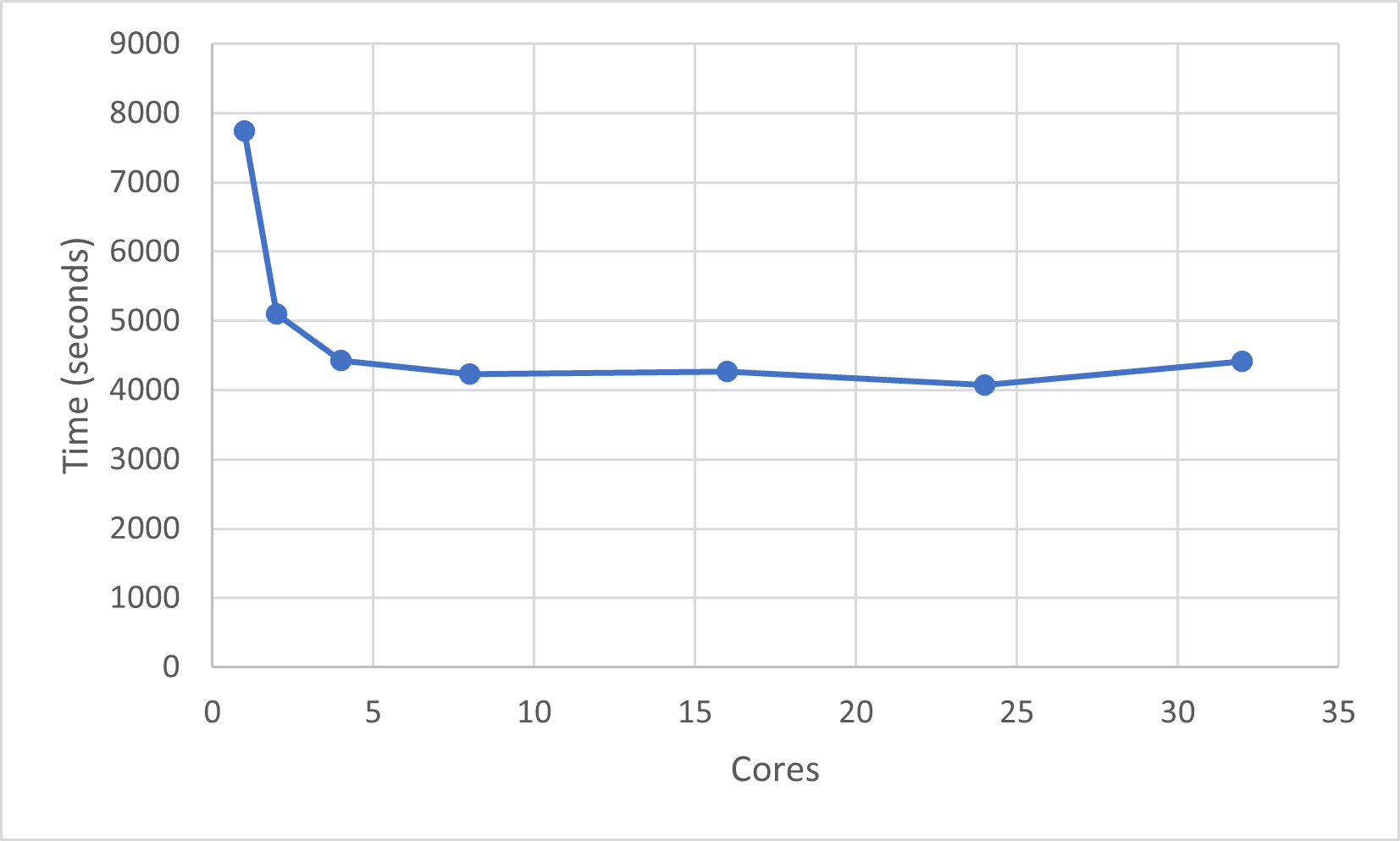}
    \caption{Execution time for the CH07 protocol verification a search depth of 10 with (a) 3 sessions and (b) 4 sessions.}
    \label{fig:Ch07-chart}
\end{figure}

\begin{figure}
        \includegraphics[width=0.5\linewidth]{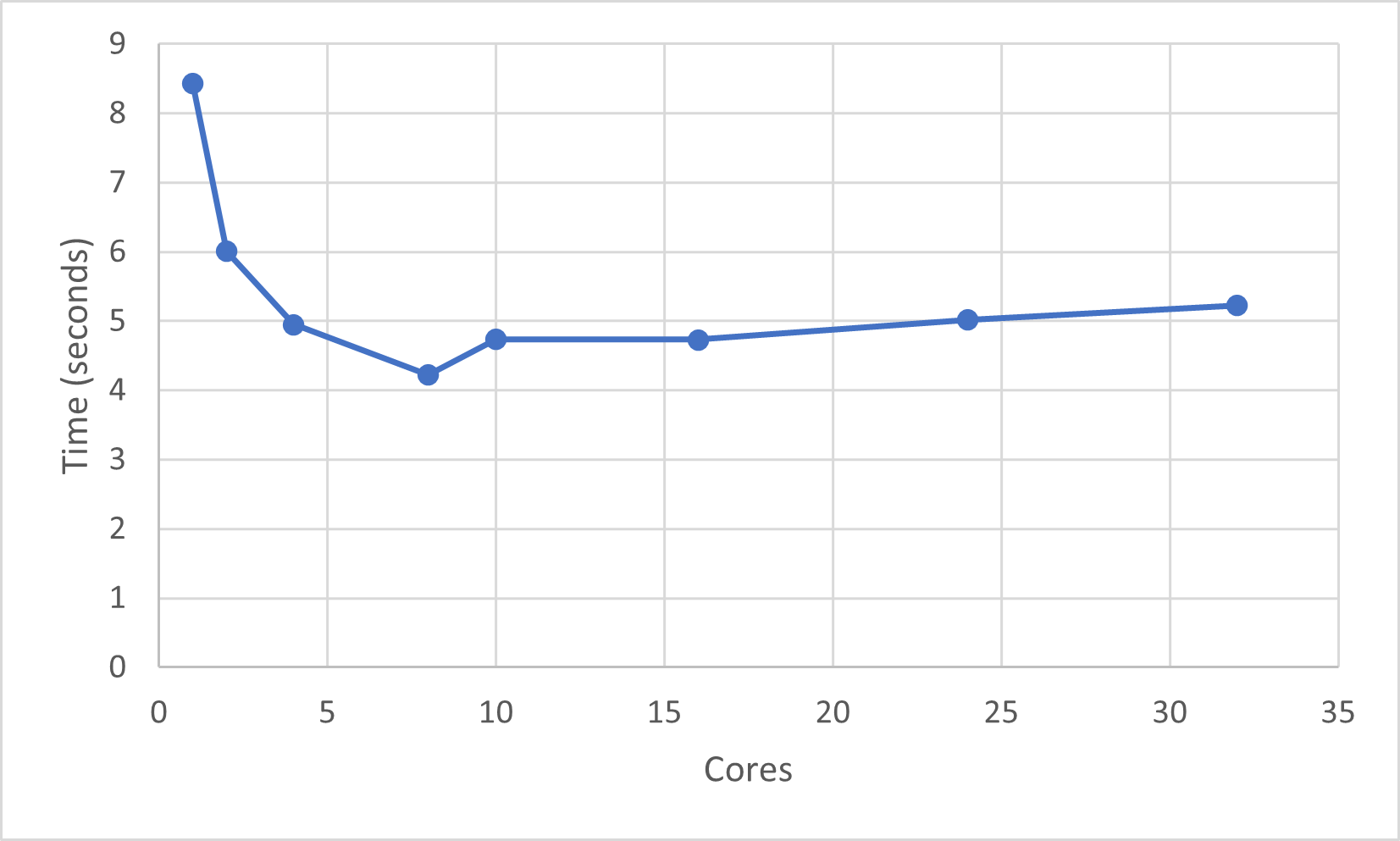}
         \includegraphics[width=0.5\linewidth]{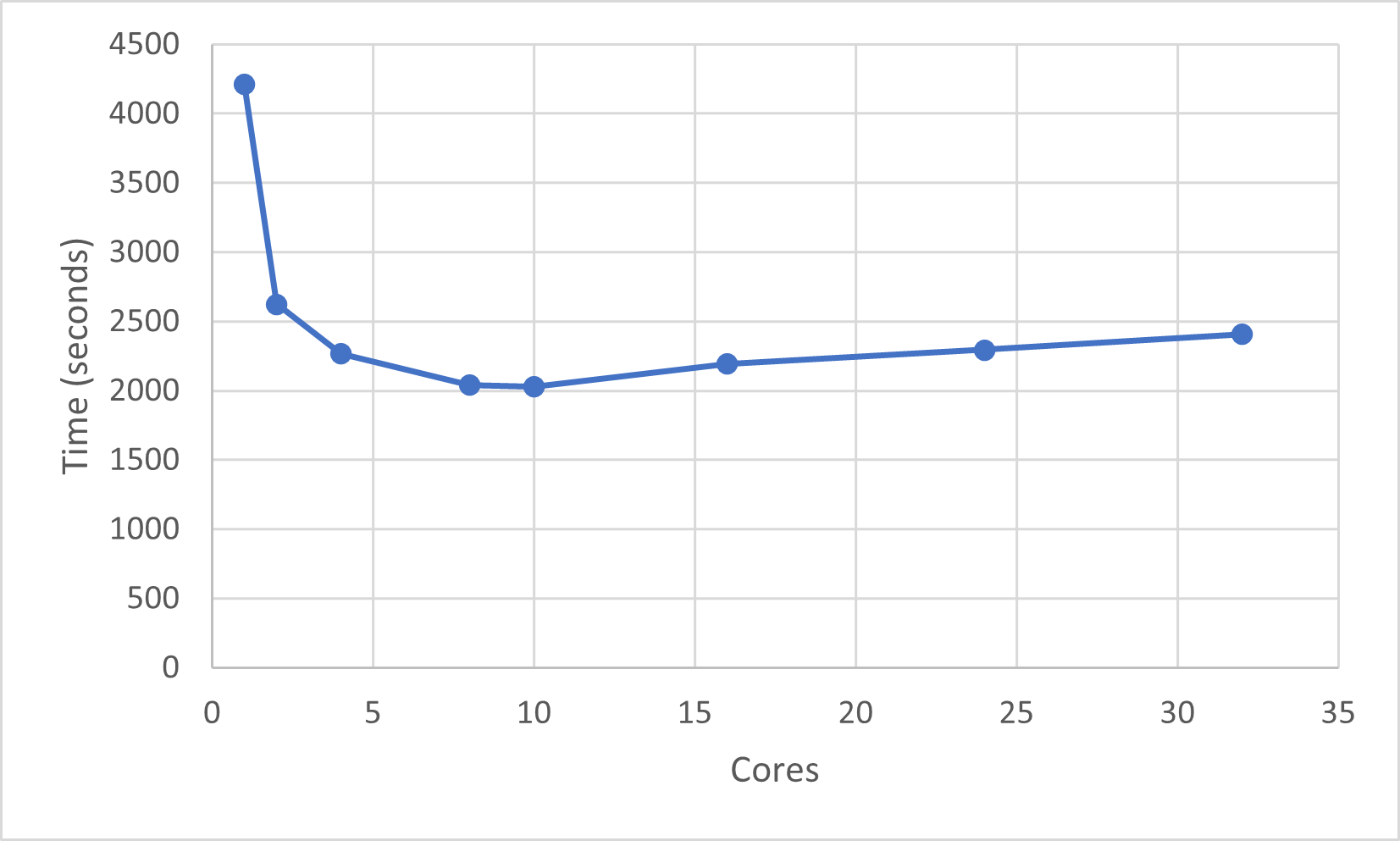}
    \caption{Execution time for the LAK protocol verification for a search depth of 10 with (a) 2 sessions  and (b) 3 sessions}
    \label{fig:LAK2-chart}
\end{figure}

We ran several protocols which require custom algebraic theories.
The execution times vs.
the number of cores for the IKA, SRA Three Pass and Salary Sum protocols from \cite{CortierDL06} and the LAK and CH07 protocols from \cite{DeursenR08} are shown in Figures
\ref{fig:IKA1-chart} to \ref{fig:LAK2-chart}. For all of the protocols, we attempted to run them with
2, 3 and 4 sessions. However, for some cases, the verification did not terminate for over 24 hrs with
 higher numbers of sessions, so only lower numbers are shown here, e.g. the Salary Sum protocol could not be verified with more than 2 sessions. All of the protocols have known attacks, but for the SRA and LAK
 protocols, OFMC does not find any attacks (see Section \ref{attacks}).

For all the models, using multiple cores produced an improvement in execution time compared to a single
core. Similar to what we observed in the case of Dolev-Yao intruder models, as the number of cores increased, the execution times generally decreased. However, after around 16 cores, the execution times started to increase again, as the garbage collection overheads increased, although the overall execution time still remained below the original time
for a single core. 

Another interesting point to observe is that for each particular protocol, e.g. the LAK protocol in Fig. \ref{fig:LAK2-chart}, the shape of the graphs are identical for different numbers
of sessions, although with significantly different execution times.


\subsection{Attacks Not Found by OFMC}\label{attacks}

Extending OFMC to improve support for custom theories has revealed some limitations of OFMC.
There are some attacks related to algebraic operators that OFMC does not find. 

The LAK protocol is an example where OFMC does not find an attack which requires the properties
of the algebraic operators. There is an attack on the LAK protocol where the attacker utilises the 
properties of XOR to obtain some secret information \cite{DeursenR08}. However, OFMC reports no attacks for this protocol.
This may be due to the fact that OFMC uses a bound on the message term depth to enable the verification
of algebraic operators to be decidable \cite{BasinMV05}. Unfortunately, this means that for several of  
the protocols we tested, OFMC could not find the known attacks.
   
\section{Profiling Tamarin's parallelisation}
\label{sec:tamarin}
In this section, we show our preliminary tests on Tamarin parallelisation, using selected example specifications included in the Tamarin distribution. 
The following experiments were performed on an Ubuntu 18.04 server, running on a machine with  192 cores Intel(R) Xeon(R) Gold 6252 and 187GB RAM. 
The Tamarin version we were using was version 1.7.1, from the "develop" branch of the git repository (git version: 98058d7d6282edfd087aefa97c3a3baa6a34ac63).

Table~\ref{tbl:tamarin1} shows the elapsed time for the following benchmark problems from Tamarin distribution:\\

\begin{tabular}{rl}
    Chen\_Kudla: & \scalebox{1}{\lstinline{examples/ake/bilinear/Chen_Kudla.spthy}}\\
    UM\_three\_pass: & \scalebox{1}{\lstinline{examples/ake/dh/UM_three_pass.spthy}}\\
    commitment-protocol: & \scalebox{1}{\lstinline{examples/csf17/commitment-protocol.spthy}}\\
    gcm: &  \scalebox{1}{\lstinline{examples/csf19-wrapping/gcm.spthy}}\\
    arpki-NoThreeUntrusted: & \scalebox{1}{\lstinline{examples/ccs14/arpki-NoThreeUntrusted.spthy}}\\
\end{tabular}

\vskip 11pt

\begin{figure}[t]
    \begin{tabular}{c@{\qquad}c}
    \includegraphics[width=0.5\linewidth]{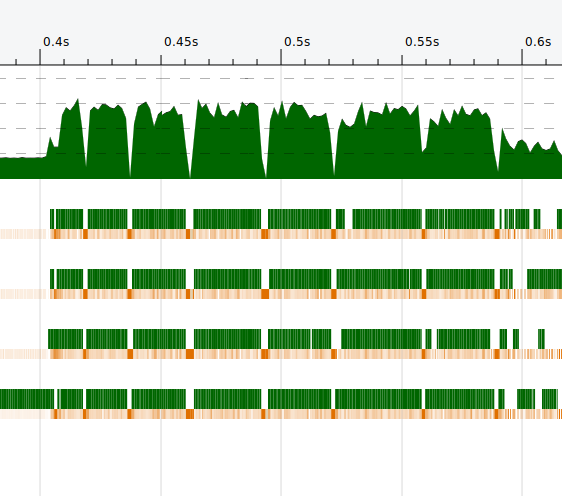} & 
    \includegraphics[width=0.5\linewidth]{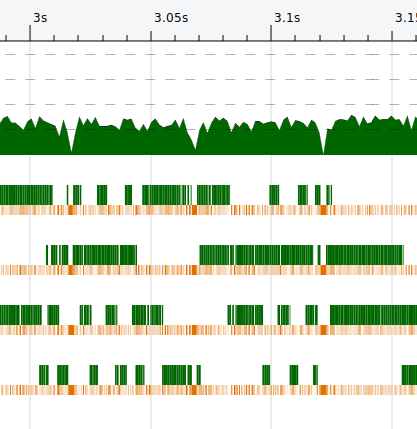} \\
    (a) & (b) \\
    \end{tabular}
    \caption{Runtime profiling of Chen\_Kudla benchmark}
    \label{fig:tamarin1}
\end{figure}

It seems that generally, there is little improvement in the elapsed time beyond 2 cores. 
Figure~\ref{fig:tamarin1} shows some fragments of the thread profiling for the \lstinline{Chen_Kudla} benchmark, running on 4 cores. 
Figure~\ref{fig:tamarin1}(a) shows an example situation where the workload was distributed evenly among the four cores. This occurred for a very short duration (less than a second). The rest of the runtime profile resembles more the situation shown in Figure~\ref{fig:tamarin1}(b) where only two cores were actively used for the actual computation (the rest was either spent on garbage collection or idle). This same pattern was observed across the benchmarks we tested (except for the \lstinline{arpki-NoThereUntrusted}, for which we did not generate a runtime profile due to the size of the log that it generates). 

For the last benchmark, we did observe a relatively more significant improvement in the elapsed time; however the gain does not seem to scale with the number of cores used. It is unclear at the moment where the bottleneck for parallelisation lies; this is left for future work. 

\begin{table}\scalebox{1}{
    \begin{tabular}{r|r|r|r|r|r}
    Benchmark & 1 core & 2 cores & 4 cores & 8 cores & 16 cores\\
    \hline
    \lstinline{Chen_Kudla} &  39.270s & 31.380s & 32.080s & 32.040s & 35.060s \\
    \lstinline{UM_three_pass} & 95.310s & 82.720s & 84.520s & 83.800s &  83.260s \\
    \lstinline{commitment-protocol} & 100.750s & 73.700s & 64.820s & 62.080s & 62.750s \\
    \lstinline{gcm} & 89.270s & 81.620s & 59.280s  & 66.940s  & 78.890s  \\
    \lstinline{arpki-NoThreeUntrusted} & 6293.050s & 4669.480s & 3758.720s & 3413.110s & 3686.980s \\
    \end{tabular}}
    \vspace{2mm}
    \caption{Total elapsed time}
    \label{tbl:tamarin1}
\end{table}

\end{document}